\RequirePackage{fix-cm}
\documentclass{svjour3}                     
\smartqed  
\usepackage{graphicx}
\usepackage{algorithm,algorithmicx,algpseudocode,amsmath}
\usepackage{soul} 
\usepackage[compatibility=false]{caption}
\usepackage{graphicx}
\usepackage{subcaption}
\usepackage{color}
 1

\begin{document}

\title{\textcolor{red}{Manuscript accepted for publication in Springer Wireless Networks} \\ \\
Improving DL-MU-MIMO Performance in IEEE 802.11ac Networks through Decoupled Scheduling}

\author{Katarzyna Kosek-Szott}


\institute{K. Kosek-Szott \at
              AGH University of Science and Technology, al. Mickiewicza 30, 30-059 Krakow, Poland\\
              \email{kosek@kt.agh.edu.pl}           
}

\date{Received: date / Accepted: date}

\maketitle

\begin{abstract}
The IEEE 802.11ac standard introduces new downlink multi-user MIMO (DL-MU-MIMO) transmissions to up to four users in order to increase spatial reuse in wireless local area networks (WLANs). We argue that even better WLAN performance can be achieved by slightly modifying the DL-MU-MIMO scheduling. To this end we propose a new queuing mechanism based on the decoupling of EDCA and DL-MU-MIMO scheduling (DEMS) to avoid head-of-line blocking. We show that DEMS outperforms traditional 802.11ac scheduling based on first in, first out transmission queues. The improvement is shown in terms throughput achieved with: (a) more efficient channel usage, (b) increased probability of transmission of high priority traffic, and (c) decreased competition between frames destined to distinct users. 
\end{abstract}

\section{Introduction}
\label{intro}
{N}{ext} generation Wi-Fi networks are expected to support transmission speeds of several Gb/s and become a cable-replacement technology. There are also great expectations to use Wi-Fi connections for cellular data offloading since the use of mobile devices and the popularity of bandwidth consuming applications is growing.  ``Global mobile data traffic grew by $74\%$ in 2015 and it is expected to grow at a compound annual growth rate of $53\%$ from 2015 to 2020, reaching 30.6 exabytes per month in 2020" \cite{cisco}. To meet these expectations, IEEE~802.11 is currently being extended to support very high throughput. The 802.11ac and 802.11ad amendments have been released in 2013 and 2012, respectively, and the TGax task group for high efficiency WLAN was started in May 2013. 

\begin{figure}[htb]
\centering
\includegraphics[width= 0.4\columnwidth]{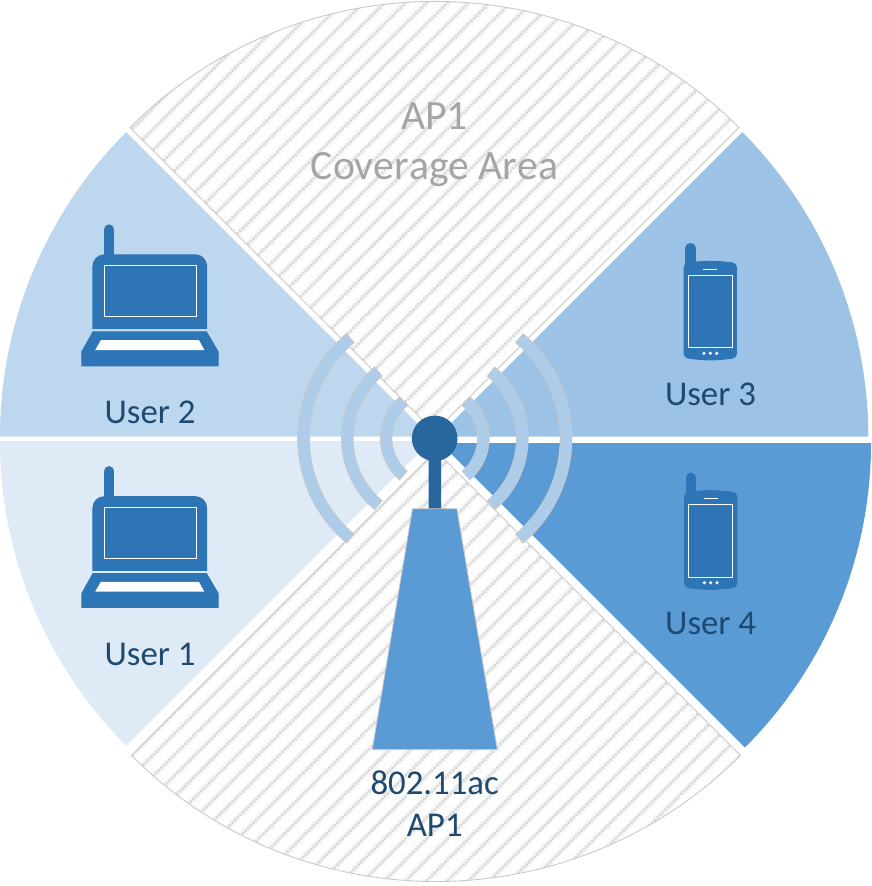}
\caption{Increased spatial reuse achieved with 802.11ac downlink MU MIMO transmissions. Up to four distinct users are considered in 802.11ac.}
\label{fig_MU-MIMO}
\end{figure}

Among the new features offered by 802.11, support for downlink (DL) multi-user multiple input multiple output (MU-MIMO) 802.11ac transmissions has been given considerable attention. The new feature allows an AP equipped with \textit{m} antennas to simultaneously serve up to \textit{m} distinct users from a group of users associated with it. As a result, spatial reuse is increased and the efficiency of Wi-Fi may be considerably improved, especially when deployed on a small-cell basis, e.g., in home or office environments.  In 802.11ac \textit{up to four distinct users} can be simultaneously served over the same wireless channel (Fig.~\ref{fig_MU-MIMO}) in the downlink direction. The user selection mechanisms were not defined in the 802.11ac standard and are out of the scope of this paper. We refer the reader to \cite{Han2016,Ma2012,Sur2016,Jeong2015} for more information about the recent advances in MU-MIMO user selection algorithms.

With DL-MU-MIMO, new quality of service (QoS) provisioning mechanisms can also be applied, especially since 802.11ac does not put many constraints on the design of the queuing mechanism in the access point (AP) \cite{gast2013}.
Additionally, 802.11ac defines a new enhanced distributed channel access (EDCA) transmission opportunity (TXOP) sharing method based on DL-MU-MIMO capabilities  \cite{ieee2013ac}. It allows the parallel transmission of frames from different access categories (ACs). The AC which wins the EDCA contention is called the \textit{primary AC}, all other ACs are called \textit{secondary ACs}. TXOP sharing is allowed when there are enough resources to transmit secondary traffic without exceeding the time required to transmit primary traffic. 

\begin{figure}[htb]
\centering
\includegraphics[width= 0.4\columnwidth]{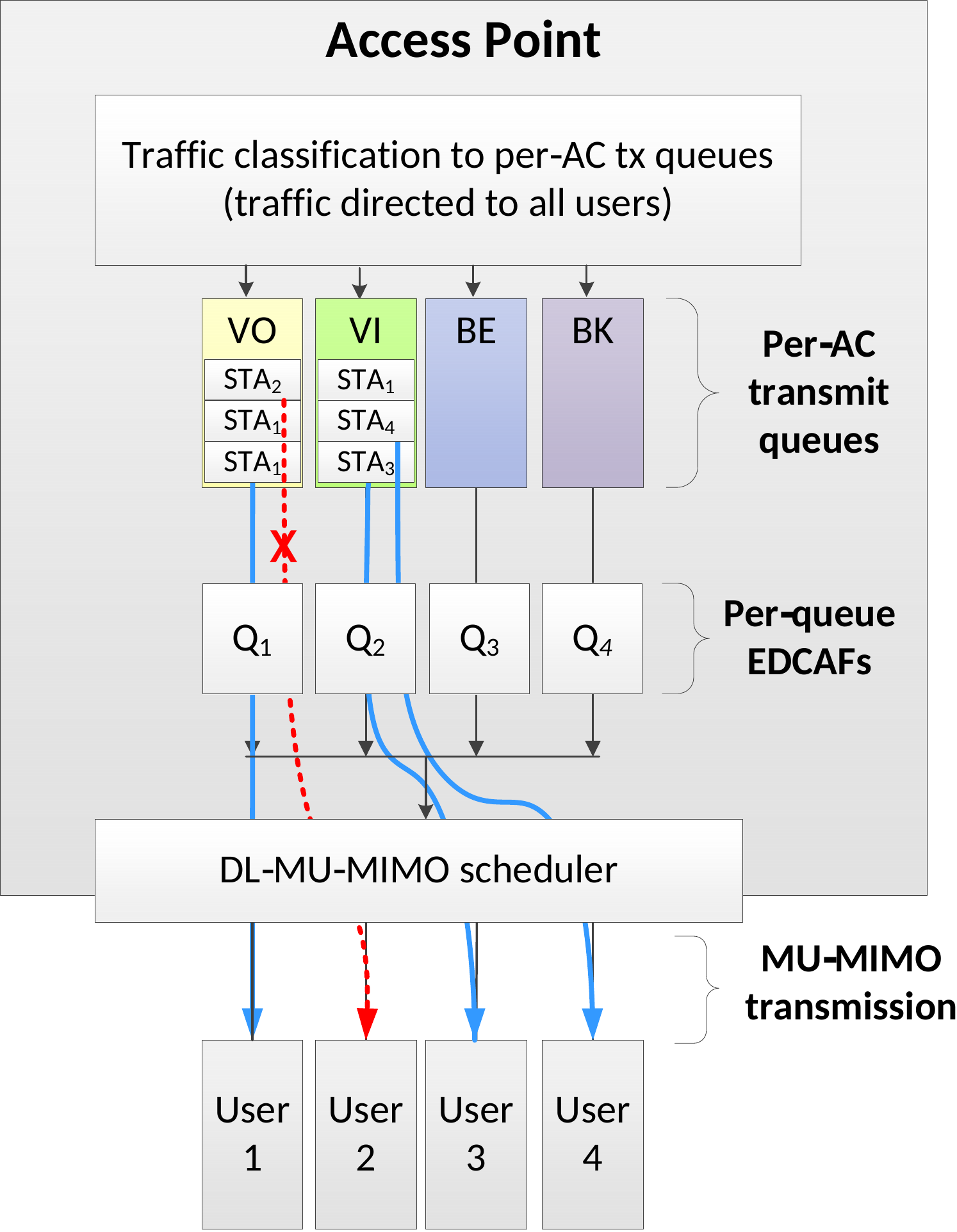}
\caption{Example of 802.11ac DL-MU-MIMO FIFO scheduling. DL-MU-MIMO simultaneous transmissions are limited to four users. With FIFO scheduling the voice (VO) frame destined to User~2 is blocked by the VO frame destined  to User~1.
}
\label{fig_old}
\end{figure}

Traditionally, in the 802.11 standard the selection of frames from the transmission (tx) queues is done in the first in, first out (FIFO) manner. Therefore, there is the possibility of wasting bandwidth due to head-of-line (HOL) frame blocking. In Fig.~\ref{fig_old}, a frame destined to User 2 is blocked by a frame destined to User 1 since a DL-MU-MIMO transmission can only contain frames destined to distinct users. 

When we consider an 802.11ac AP with single type flows (of a single AC) and homogeneous traffic (i.e., the arrival rate of flows destined to each user associated with the AP is the same), the probability of HOL blocking of the transmit queue ($p^{HOL}_{blk}$) can be easily calculated as
\begin{align}
  p^{HOL}_{blk}=\left\{
  \begin{array}{@{}ll@{}}
    1-\frac{n_u!}{(n_u-n_s)!n_u^{n_s}}, & \text{if}\ n_u\ge n_s \\
    0, & \text{otherwise,}
  \end{array}\right.
\end{align} 
where $n_u$ is the number of the served users, $n_s$ is the number of available spatial streams, $p_{opt}=\frac{n_u!}{(n_u-n_s)!n_u^{n_s}!}$ is the probability of having frames destined to $n_s$ distinct users in the optimal order (i.e., without HOL blocking), and $p_{blk}^{HOL}=0$ means that users are served without HOL blocking. Additionally, $p_{opt}$ is calculated as a fraction of the number of optimal orders of frames destined to $n_u$ distinct users having $n_s$ spatial streams $\bigg(\frac{n_u!}{(n_u-n_s)!}\bigg)$ to the total number of possible frame orders having $n_s$ spatial streams ($n_u^{n_s}$). 

\begin{figure}[htb]
\centering
\includegraphics[width= 0.5\columnwidth]{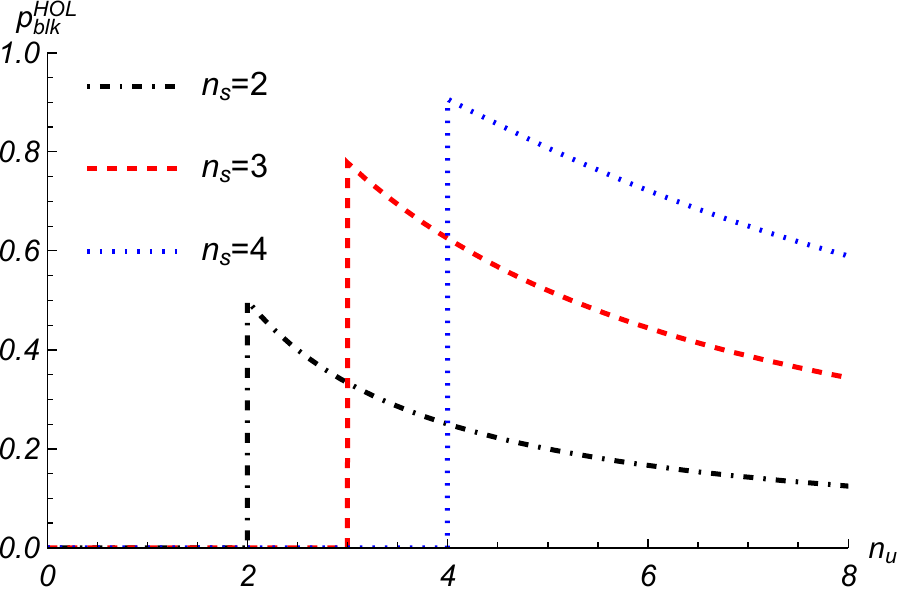}
\caption{Head of line blocking probability in an AP for a single AC and homogeneous traffic. In 802.11ac and 802.11ax up to, respectively, four and eight users can be served simultaneously.}
\label{hol}
\end{figure}

Fig.~\ref{hol} illustrates the HOL blocking probability for a single queue with variable $n_u$ and $n_s=\{2,3,4\}$.  For a given $n_u$, $p^{HOL}_{blk}$ increases with the increasing $n_s$, since the probability $p_{opt}$ decreases, i.e., the probability of transmitting to $n_s$ distinct users decreases. Additionally, for a given $n_s$, $p^{HOL}_{blk}$ decreases with the increasing number of the served users $n_u$ because the probability $p_{opt}$ increases, i.e., the probability of having  a sequence of $n_s$ frames destined to $n_s$ distinct users in a queue increases.

In order to leverage the opportunity of simultaneous MU transmissions, sophisticated queuing mechanisms can be used \cite{Wang}. This approach, however, unnecessarily increases the complexity of wireless chipsets. 

Therefore, in this paper we propose a new queuing mechanism based on the decoupling of EDCA and DL-MU-MIMO scheduling, to which we refer as DEMS queuing. 
The new queuing mechanism improves QoS provisioning in comparison to the traditional 802.11ac scheduling with FIFO queues, since it eliminates the problem of HOL blocking by introducing \emph{per-user per-access category virtual queues}.

The importance of separating queues to prevent HOL blocking when designing a downlink scheduler in DL-MU-MIMO systems was already considered in IEEE 802.16 \cite{Chakchai09} and LTE \cite{Ouyang} networks. Therefore, it confirms our observation that with applying appropriate scheduling methods (e.g., the proposed DEMS), better performance can also be achieved in future 802.11 networks. To the authors' best knowledge, this is the first paper which presents the  performance results of this feature taking into account the specifics of 802.11 networks.

The rest of the paper is organized as follows. In Section~\ref{relwork} related work is presented. Section~\ref{DEMSq} presents the details of DEMS queuing. Section~\ref{examples} contains operational examples illustrating how DEMS queuing improves QoS provisioning in 802.11ac networks. Simulation results are presented in Section~\ref{sim_res}: they compare the upper bound throughput values achieved with DEMS and traditional 802.11ac scheduling with FIFO queues. Concluding remarks are presented in Section~\ref{concl}. 

\section{Related Work}
\label{relwork}
There are several papers in the literature  devoted to DL-MU-MIMO transmissions in 802.11 networks and their implications on the resource sharing and the data link layer of the OSI model \cite{bellalta2012performance,cha2012performance,Cheng14,chung2013mpdu,valls2013proportional,Chunhui}. 
There is also a recent survey of MU-MIMO medium access control (MAC) protocols for IEEE 802.11 \cite{Liao14} in which the authors describe requirements for designing MU-MIMO MAC protocols and present different examples of such protocols proposed in the literature. 
These papers, however, consider DL-MU-MIMO scheduling with FIFO queues.

The problem of HOL blocking of FIFO queues is not new. In the literature there are many papers devoted to this problem for wired networks (e.g., \cite{Anderson,Zhang1}). They highlight the low efficiency and unfairness problems occurring in case of FIFO queues. As a remedy to the HOL blocking problem several different approaches have been proposed, however, the most straightforward solution with the largest performance gains is implementing virtual queues \cite{Anderson}.

Additionally, papers \cite{Chakchai09} and \cite{Ouyang} describe the idea of separating queues for IEEE 802.16 and LTE networks, respectively. In both works the authors highlight that by implementing per-user or per-connection virtual queues an optimal system performance can be achieved, since the problematic HOL blocking can be avoided. 

The proposed DEMS solution is partially similar to \cite{ruckus}. The Ruckus company designed a SmartCast mechanism which was implemented in legacy 802.11 APs. It implements per-user software queues as contrary to the traditionally used per-traffic class queues in order to avoid HOL blocking. Strict priority scheduling across traffic classes is used and weighted round robin scheduling is used across different users. However, strict priority scheduling considered by SmartCast may lead to the starvation of low priority flows. Additionally, to the authors' best knowledge, in the literature there is no information on the integration of SmartCast with 802.11ac DL-MU-MIMO. Finally, the details of SmartCast's implementation are confidential.

Finally, the newest patch for the ath9k driver introduces novel per-station per-traffic class mac80211 intermediate software queues in order to avoid buffer bloats in case of large aggregated frames\footnote{https://patchwork.kernel.org}. Therefore, the implementation of DEMS in real devices is feasible.

\section{DEMS Queuing Overview}
\label{DEMSq}

The DEMS queuing mechanism proposed in this paper is based on the decoupling of EDCA and DL-MU-MIMO scheduling, which takes advantage of EDCA TXOP sharing without the need for implementing sophisticated queuing mechanisms.

\begin{figure}[htb]
\centering
\includegraphics[width=0.5 \columnwidth]{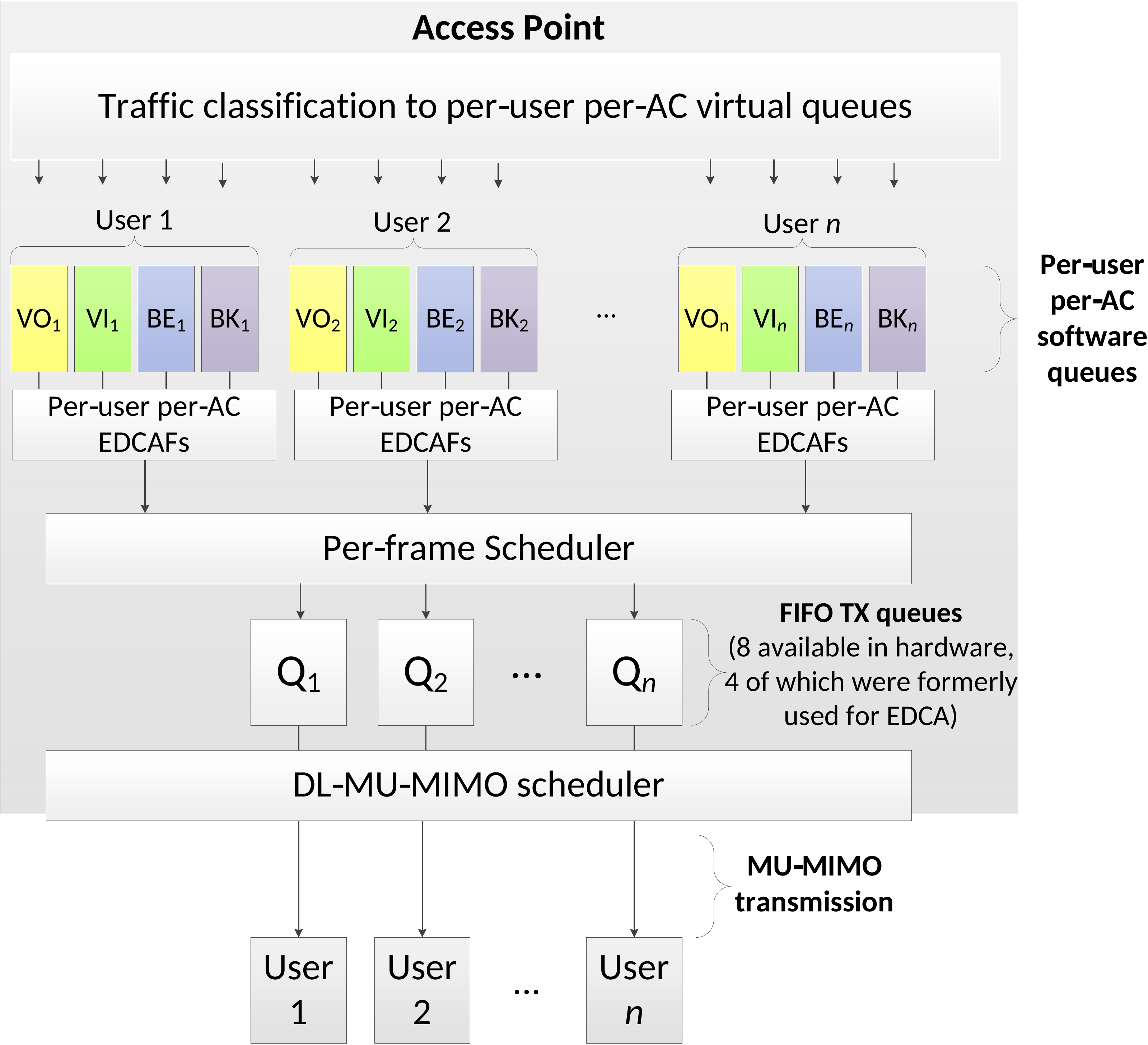}
\caption{DEMS queuing proposal. Instead of per-AC FIFO hardware tx queues, new per-user FIFO hardware tx queues are proposed. Per-user per-AC EDCAFs operate in the virtual (software) domain.}
\label{fig_new}
\end{figure}

DEMS uses four \textit{virtual FIFO class queues} for each user associated with the AP (Fig.~\ref{fig_new}). Therefore, frames arriving from the higher layers are classified to these \textit{virtual per-user per-AC queues}. The contention between ACs is organized using the standard EDCA procedure.
Afterwards, a per-frame scheduler selects frames destined to non-interfering users. The exact scheduling method is out of the scope of this paper and is left as future work. However, it is an important part of the system because it not only may improve system scalability (more users than the available number of tx queues can be served) but also may increase its flexibility (by using different scheduling algorithms, different network management goals can be achieved, e.g., airtime fairness or prioritized service of certain users). After frames are selected by the per-frame scheduler, they are put into \textit{hardware FIFO transmission queues}. 
Importantly, the IEEE 802.11ac standard limits DL-MU-MIMO transmissions to four users and it is planned that IEEE 802.11ax will support up to eight users. Additionally, eight hardware queues are available in typical commercial off-the-shell equipment. 
Therefore, as presented in Fig.~\ref{fig_new}, DEMS is open to increasing the number of simultaneous transmissions in the future.

Once the channel is ready for transmission, frames destined to non-interfering users are scheduled for a DL-MU-MIMO transmission. To ensure backward compatibility, EDCA parameters of the highest priority frame included in the DL-MU-MIMO frame dictate the channel access rules. 
Algorithm \ref{alg-DEMS} presents the proposed initial DEMS implementation in an AP.
As future work, DEMS queuing can additionally be coupled with other IEEE 802.11ac improvements proposed in the literature, e.g., frame aggregation \cite{bellalta2012performance} or frame fragmentation \cite{chung2013mpdu}, to further improve channel utilization.

\begin{algorithm}[htb]
\caption{Proposed DEMS queuing in an AP}\label{ttq}
\small
\begin{algorithmic}[3]
\Function{EDCA-based MAC scheduling}{}
\State {$N_u \gets $ Number of non-interfering receivers (users) selected by the per-frame scheduler}
\State $N_{AC} \gets \text{Number of ACs}$
\State {$\alpha_{i,j} \gets$ Frame selection probability for the $i$-th class queue of user $j$}
\State $Q_{j} \gets \text{Transmission queue for user $j$}$
\State {$f_{i,j} \gets$ HOL frame to $j$-th user from the $i$-th class queue}
\For {$i \gets 0:N_{AC}-1$ }
\For {$j \gets 1: N_u$} 
\State With probability  $\alpha_{i,j}$ move $f_{i,j}$ to $Q_j$
\EndFor
\EndFor
\State \Return{$Q_{j}$}
\EndFunction

\Function{DL-MU-MIMO scheduling}{}
\For {$j \gets 1: N_u$} 
\State Select HOL frame from $Q_j$ 
\EndFor
\State \Return DL-MU-MIMO frames
\EndFunction
\end{algorithmic}

\label{alg-DEMS}
\end{algorithm}

Importantly, DEMS queuing requires mostly software changes in the wireless devices since the FIFO transmission hardware queues remain unchanged. 
Additionally, the Ruckus SmartCast solution \cite{ruckus}, being in line with our proposed approach, proves that practical implementation of DEMS is feasible and may lead to considerable Wi-Fi network performance improvements. The implementation of DEMS will become even easier once wireless software defined networking (SDN) platforms will provide DL-MU-MIMO support \cite{Bianchi2012}. Meanwhile, in this paper, after analyzing examples of DEMS operation in different scenarios (Section \ref{examples}), we validate our concept by simulations (Section \ref{sim_res}).

\section{Operational Examples}
\label{examples}

\begin{figure}[htb]
\centering
\includegraphics[width= 0.8\columnwidth]{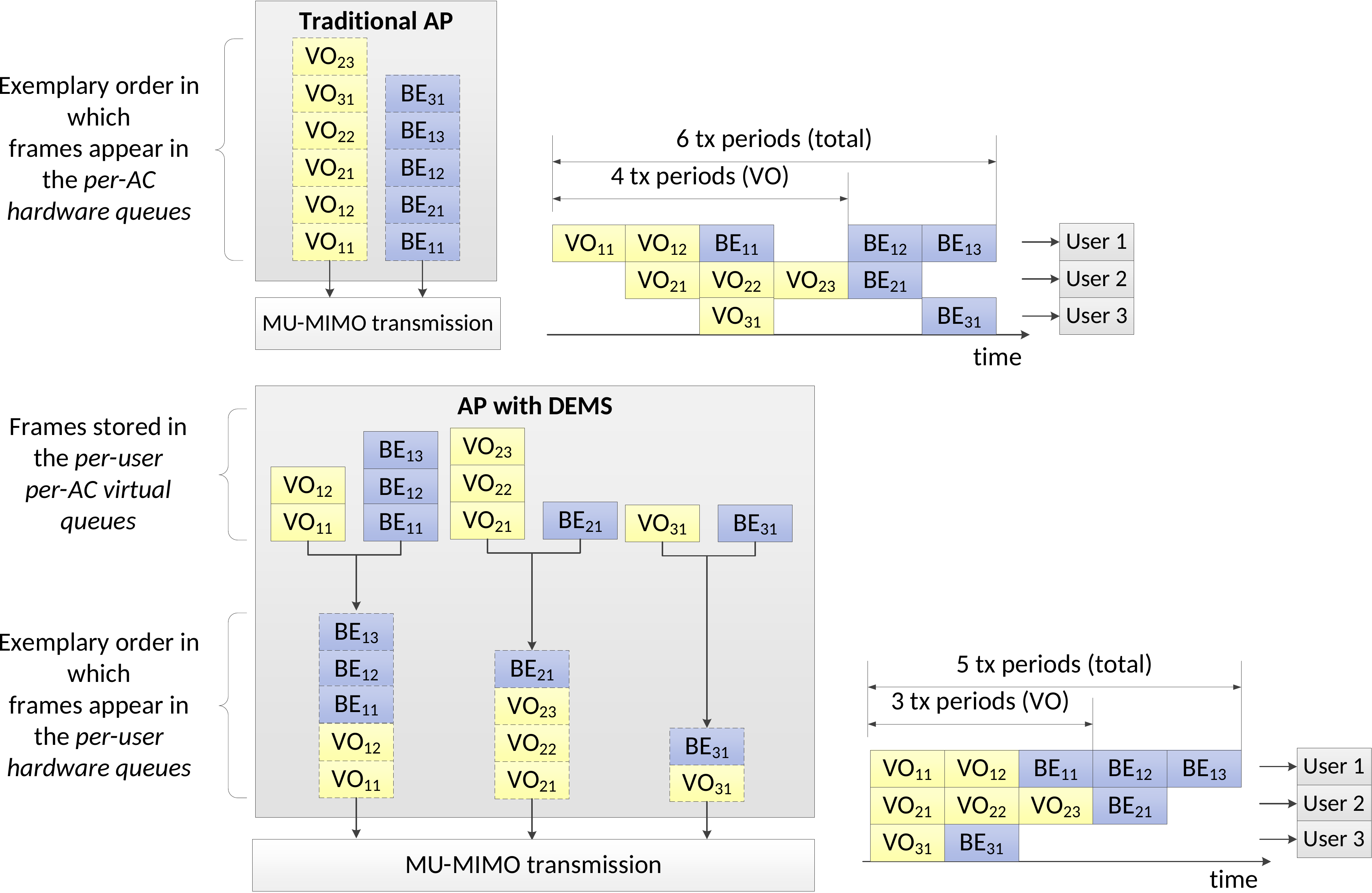}
\caption{Example of EDCA TXOP sharing for equal length frames ($l_{VO}=l_{BE}$) with traditional (top) and DEMS (bottom) queuing  under the assumption of three users.}
\label{fig_tx}
\end{figure}

\begin{figure}[htb]
\centering
\includegraphics[width= 0.8 \columnwidth]{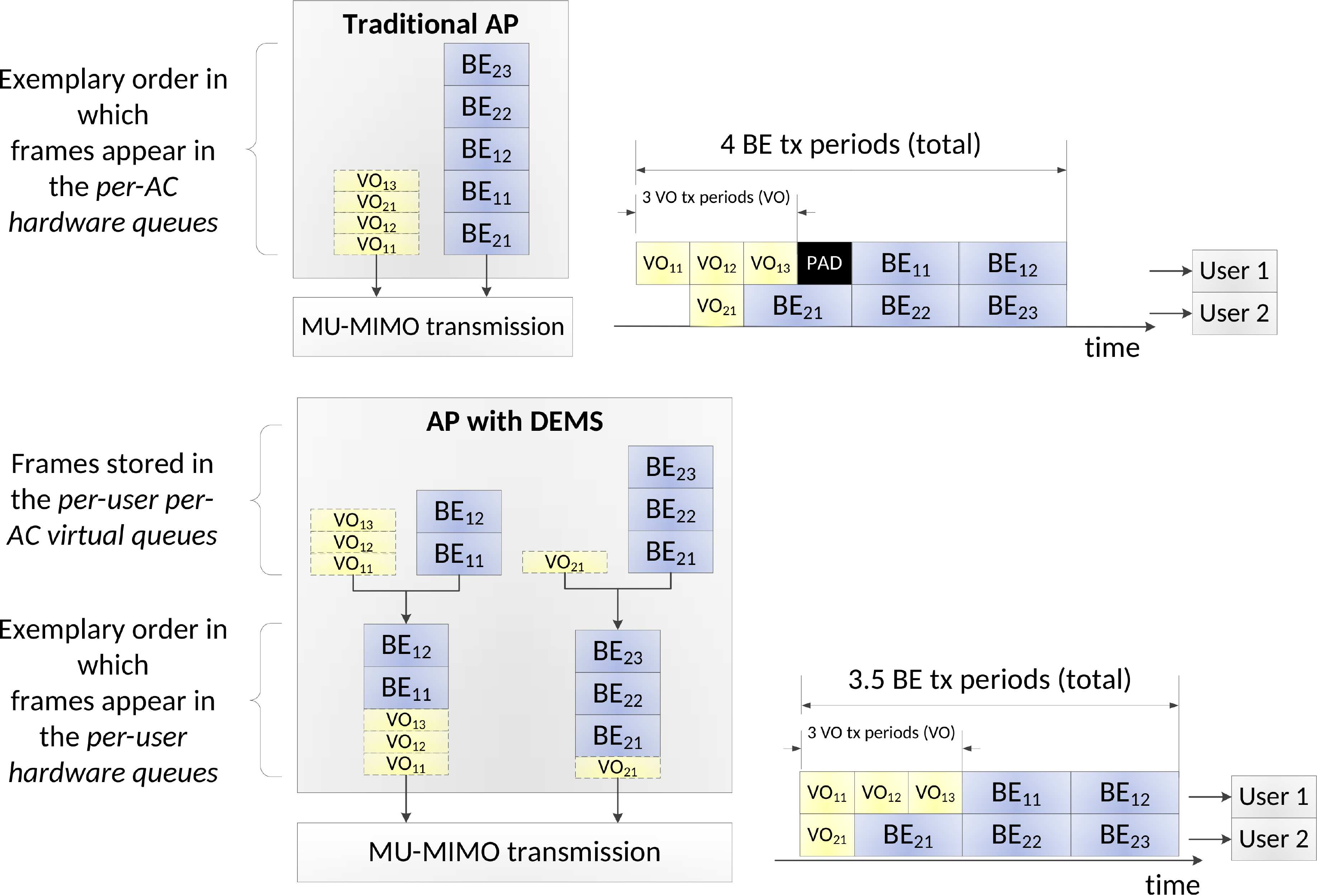}
\caption{Example of EDCA TXOP sharing for varying frame lengths $l_{VO}= 2 \times l_{BE}$: traditional (top) and DEMS queuing (bottom) under the assumption of two users.}
\label{fig_tx2}
\end{figure}

\begin{figure}[htb]
\centering
\includegraphics[width=0.9  \columnwidth]{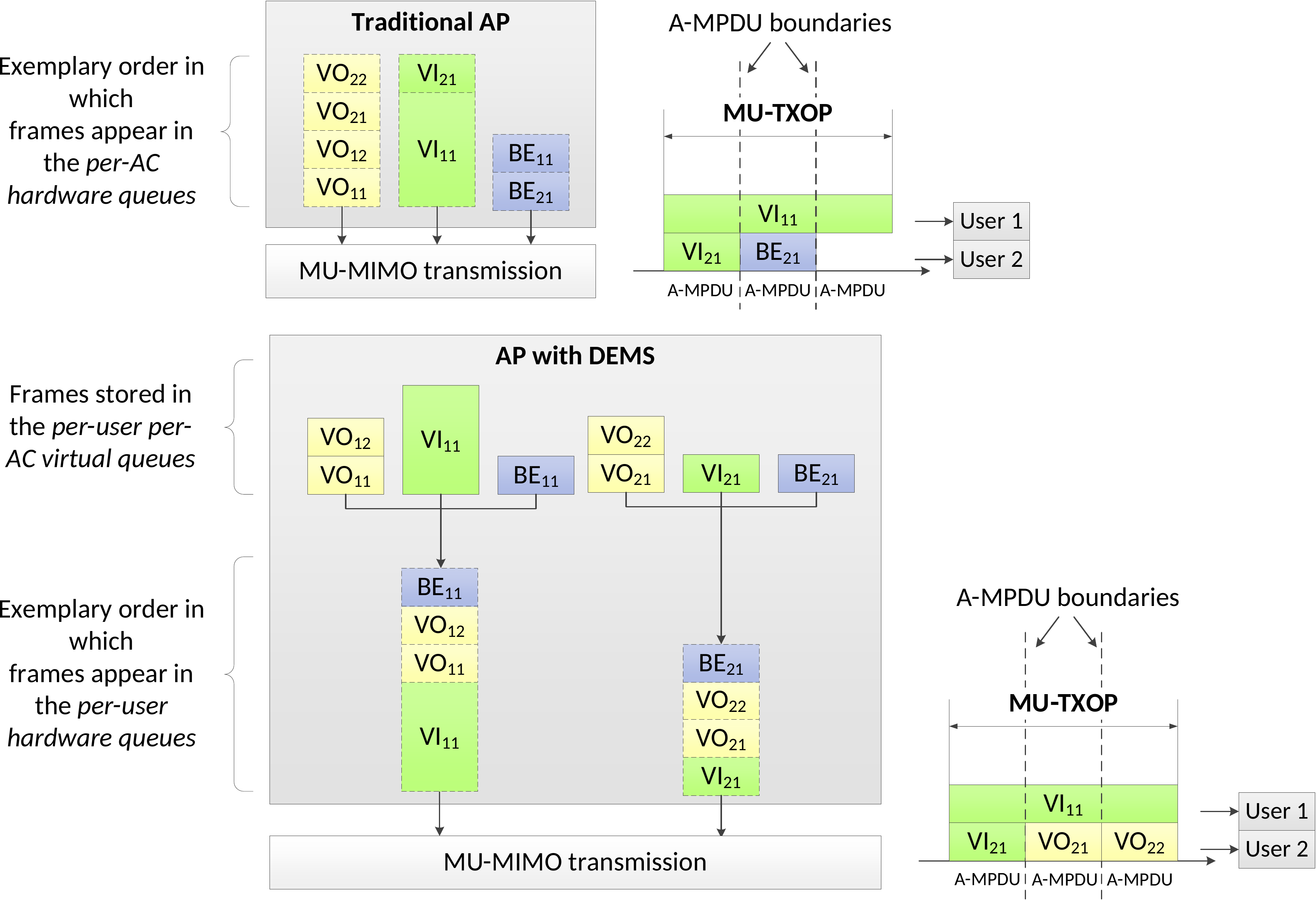}
\caption{Example of EDCA TXOP sharing with traditional (top) and DEMS queuing (bottom) under the assumption of two users.}
\label{fig_tx3}
\end{figure}

We present three operational examples to illustrate how DEMS queuing may improve QoS provisioning in 802.11ac networks.
For clarity of presentation, only frames transmitted by the AP are shown (e.g., block acknowledgments are omitted, %
for more information on the acknowledgment procedure we refer the reader to \cite{ieee2013ac}
) and selected EDCA ACs are considered (high priority voice -- VO, high priority video -- VI, and low priority best effort -- BE). In all figures DEMS is compared with the traditional scheduling for a finite number of frames to illustrate their operation. Importantly, the order in which the frames are transmitted over the wireless medium depends on per-queue EDCA scheduling for 802.11ac and per-user per-AC EDCA scheduling for DEMS. Additionally, Figs. \ref{fig_tx}--\ref{fig_tx3} describe each frame by its AC and two indexes ($x,y$), where $x$ is the destination user number and $y$ is the frame number (e.g., $VO_{12}$ is the second VO frame destined to User 1).

First, we assume equal VO and BE frame lengths ($l_{VO}=l_{BE}$) as well as PHY transmission rates ($r_{VO}=r_{BE}$). This is the basic example illustrating EDCA TXOP sharing in case of traditional and DEMS queuing and three distinct users. As shown in Fig.~\ref{fig_tx}, for 802.11ac the frames are delivered to users after six transmission periods (VO frames after four periods). Additionally, there are several empty transmission periods, e.g., between the VO$_{31}$ and BE$_{31}$ transmission because the latter frame is blocked by BE$_{21}$ and BE$_{12}$. With DEMS queuing all frames are delivered after five transmission periods (VO frames after three periods) and there are no empty tx periods between frames. 

In the second configuration we assume that VO frames are shorter than BE frames ($l_{BE}= 2 \times l_{VO}$). Additionally, it is assumed that data is transmitted to two users at the same data rate ($r_{VO}=r_{BE}$), and the AC VO (AC BE) is always the \textit{primary} (\textit{secondary}) AC. 
Both VO and BE frames are transmitted faster with the new mechanism (Fig.~\ref{fig_tx2}). Additionally, with DEMS it is possible to avoid not only HOL blocking but also padding, which results in even more efficient channel usage. 

In Fig.~\ref{fig_tx3} we assume that AC VI obtains the EDCA TXOP and shares it with VO and BE ACs. The VI$_{11}$ frame is a very high throughput (VHT) single MPDU fragmented into three A-MPDUs, as described in \cite{Chunhui}, which is destined to User 1. With DEMS it is possible to transmit high priority data (VO frames) before low priority data (BE frames) which is not possible with traditional scheduling (where VO frames destined to User 2 are blocked by the VO frame destined to User 1). Therefore, with DEMS not only channel usage efficiency but also flexibility in the order of frame transmissions can be improved.

In summary, the presented operational scenarios show that 802.11ac TXOP sharing can be used to improve QoS provisioning in future Wi-Fi networks but a sophisticated method of scheduling frames is required. Otherwise, channel resources are wasted and/or low priority traffic precedes high priority transmissions. The following section quantifies the improvement that can be expected using DEMS queuing in comparison to traditional  802.11ac scheduling with FIFO queues.

\section{Simulation Results}
\label{sim_res}

\begin{figure*}[htb]
\centering
\includegraphics[width= \columnwidth]{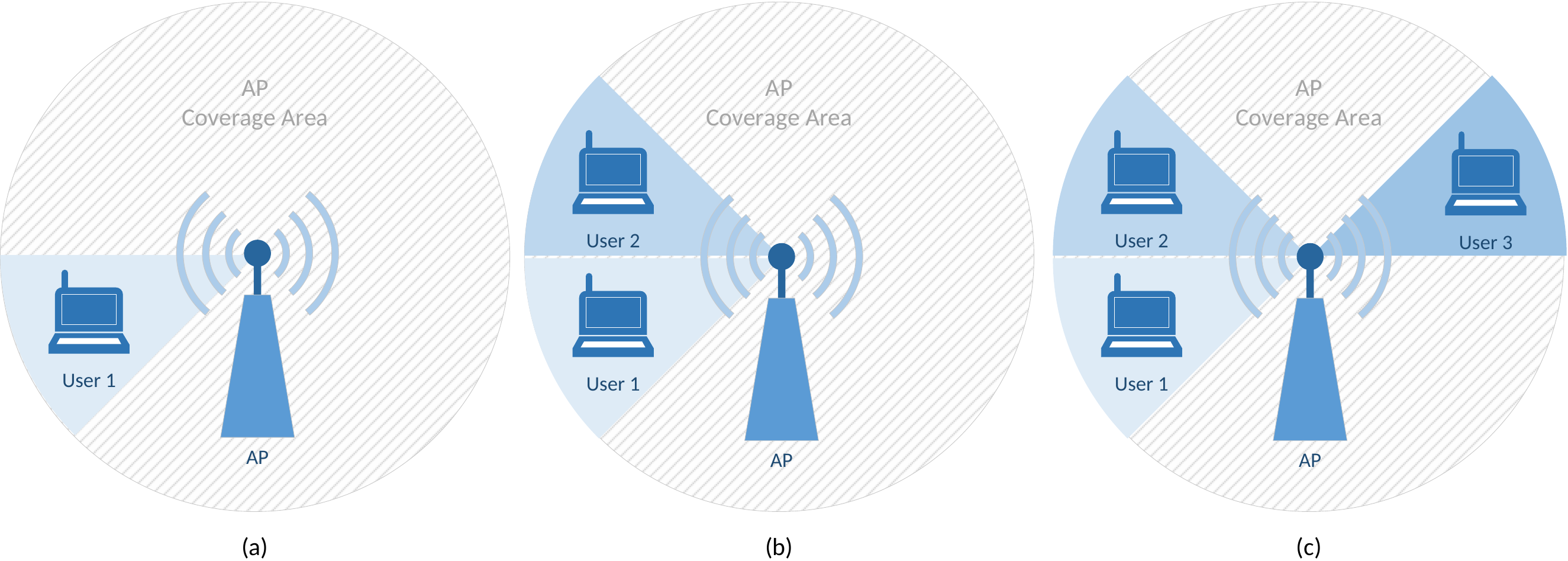}
\caption{Tested network topologies: DL-MU-MIMO transmissions with (a) a single user, (b) two users, (c) three users.}
\label{fig_scenarios}
\end{figure*}

To validate our concept, we have implemented the traditional 802.11ac scheduling with FIFO queues and the proposed DEMS queuing in Matlab. 
In our tests, the network consisted of an AP equipped with three antennas (i.e., a typical 802.11ac AP currently available on the market) and one, two, or three DL-MU-MIMO capable users, each equipped with a single antenna (as in available 802.11ac smartphones) (cf. Fig.~\ref{fig_scenarios}). 
The AP transmitted VO and BE traffic. Additionally, since we consider DL-MU-MIMO transmissions, there were no collisions at the PHY layer. In the simulations, we assumed that the frame size ($l$) and data rate ($r$) are constant and equal for all users. 
The simulation time was set to $500 \times l/r$ and 15 runs were performed for each simulation point.

The probability of selecting a VO frame from its class queue ($\alpha$), which is related to per-user MAC scheduling (cf. Fig.~\ref{fig_new} and Algorithm \ref{alg-DEMS}), varied from $\alpha_{min}=0.5$ to $\alpha_{max}=1$ (with a step of 0.02) for each user. 
The probability of scheduling an AC BE frame was set to $1-\alpha$. Therefore, the setting of $\alpha=1$ means strict priority scheduling, i.e., the higher priority frame will be always selected before the low priority one. This last setting is similar to what was proposed in SmartCast by Ruckus \cite{ruckus}. The setting of $\alpha=0.5$ means no prioritization in traffic scheduling.

The normalized amount of traffic $L$ destined to each user was set to:
\begin{enumerate}
\item \textbf{Single user case}: $L_u=\beta$, where $\beta=1$.
\item \textbf{Two user case}: $L_{User 1}=\beta$ and $L_{User 2}=1-\beta$ with $\beta$ varying from $\beta_{min}=0.05$ to $\beta_{max}=0.5$. For $\beta=0.5$ the traffic destined to both users is uniformly distributed. 
\item \textbf{Three user case}: $L_{User 1}=\beta$, $L_{User 2}=1/3$, and $L_{User 3}=1-(1/3+\beta)$ with $\beta$ varying from $\beta_{min}=0.05$ to $\beta_{max}=1/3$. For $\beta=1/3$ the traffic destined to all three users is uniformly distributed.
\end{enumerate}
Additionally, infinite DEMS virtual per-user per-AC queues and infinite 802.11ac per-AC queues were considered to simplify the analysis. The assumption of finite queues would only increase the advantage of DEMS over 802.11ac since more high priority frames would have been dropped from the 802.11ac AP high priority queue as a result of HOL blocking and the increased queuing delays would have been observed. Finally,  we assumed saturated network conditions and a perfect radio channel. The perfect channel assumption does not impact the qualitative differences in the operation of IEEE 802.11ac and DEMS, since poor channel conditions would impact both approaches in a similar way, i.e., the upper bounds of their performance  would be lowered.

\subsection{Metrics}
We compare DEMS with traditional 802.11ac FIFO queuing using the following metrics. 

\textit{Throughput change}, i.e., the measure of throughput change achieved with DEMS in comparison to 802.11ac\footnote{Positive throughput changes also mean lower queuing delays (frames spend less time in queues), since more frames are transmitted at the same time. Negative throughput changes mean longer queuing delays.}. It can be calculated for varying values of $\alpha$ and $\beta$ separately for the $i$-th AC as:
\begin{equation}
\label{thr_ch}
\small
T_{change}[i](\alpha,\beta)=
\left(\frac{c_{\text{DEMS}}[i](\alpha,\beta)}{c_{\text{802.11ac}}[i](\alpha,\beta)} -1 \right) \times 100\%,
\end{equation}
where $c_{\text{802.11ac}}$ and $c_{\text{DEMS}}$ are counters for traditional scheduling and DEMS, respectively. In our simulations these counters represent the average number of frame transmissions of the $i$-th AC destined to all users calculated over 500 transmission periods (i.e., $500 \times l/r$) and are therefore a measure of the achieved throughput. 

\textit{Average $i$-th AC throughput $T_{avg}[i]$ as a function of $\beta$}, i.e., the measure of throughput for a varying traffic distribution among 802.11ac users (i.e., DL-MU transmission receivers), calculated as:
\begin{equation}
\label{t_avg}
T_{avg}[i](\beta_j)= 
\frac{\sum\limits_{\gamma=0}^{n_\alpha-1} c[i](\alpha_{min}+\gamma  \frac{\alpha_{max}-\alpha_{min}}{n_{\alpha}-1},\beta_j)}{n_{\alpha}},
\end{equation}
where $i$ represents the AC, $j$ is the step counter for the $\beta$ parameter \big(in our simulations $j \in \{0, 1, ..., n_{\alpha}-1\}$\big), $\beta_j$ is equal to $\beta_{min}+j \frac{\beta_{max}-\beta_{min}}{n_{\alpha}-1}$, $n_{\alpha}$ is the number of steps for the $\alpha$ parameter (in our simulations it was set to 25), and $c$ represents $c_{DEMS}$ for DEMS and $c_{802.11ac}$ for 802.11ac.

\textit{Average throughput change $T_{change}^{avg}$ for the $i$-th AC as a function of $\beta$}, i.e., the measure of throughput change achieved with DEMS in comparison to 802.11ac for the varying traffic distribution among 802.11ac users, calculated as:
\begin{equation}
\label{t_ch_avg}
T_{change}^{avg}[i](\beta_j)= 
\frac{\sum\limits_{\gamma=0}^{n_\alpha-1} T_{change}[i](\alpha_{min}+\gamma  \frac{\alpha_{max}-\alpha_{min}}{n_\alpha-1},\beta_j)}{n_\alpha}.
\end{equation}

\textit{Average $i$-th AC throughput $T_{avg}[i]$ as a function of $\alpha$}, i.e., the measure of throughput for the varying priority of VO over BE, calculated as:
\begin{equation}
\label{t_avg-alpha}
T_{avg}[i](\alpha_j)= 
 \frac{\sum\limits_{\gamma=0}^{n_\beta-1} c[i](\alpha_j,\beta_{min}+\gamma \frac{\beta_{max}-\beta_{min}}{n_\beta-1})}{n_\beta},
\end{equation}
where $j$ is the step counter for the $\alpha$ parameter  \big(in our simulations $j \in \{0, 1, ..., n_\beta-1\}$\big), $\alpha_j$ is equal to $\alpha_{min}+j \frac{\alpha_{max}-\alpha_{min}}{n_\beta-1}$, $n_\beta$ is the number of steps for the $\beta$ parameter (in our simulations it was set to 25). 

\textit{Average throughput change $T_{change}^{avg}$ for the $i$-th AC as a function of $\alpha$}, i.e., the measure of throughput change achieved with DEMS in comparison to 802.11ac for the varying priority of VO over BE, calculated as:
\begin{equation}
\label{t_ch_avg-alpha}
 T_{change}^{avg}[i](\alpha_j)= 
 \frac{\sum\limits_{\gamma=0}^{n_\beta-1} T_{change}[i](\alpha_j,\beta_{min}+\gamma \frac{\beta_{max}-\beta_{min}}{n_\beta-1})}{n_\beta}.
\end{equation}

\begin{figure}[htb]
        \centering
                \includegraphics[width=0.65\columnwidth]{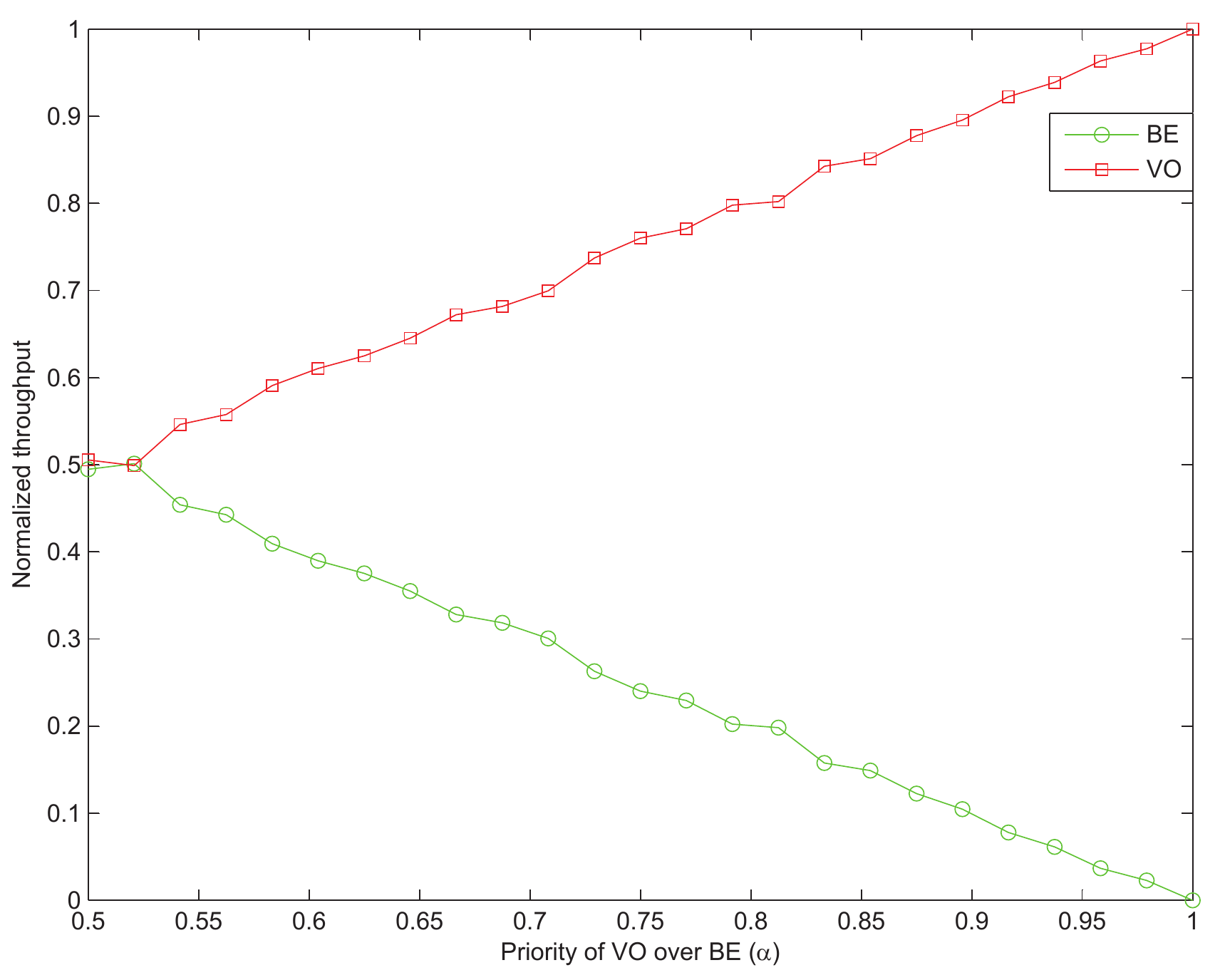}
\caption{Single user case:  average normalized VO and BE throughput. A typical value of $\alpha$ for EDCA under saturation is about $0.85$ \cite{kosek2011simple}.
}
\label{fig_thr_EDCA}
\end{figure}

\subsection{Single User Case}
In case of a single user, DEMS and 802.11ac perform exactly the same. Additionally, their behavior is similar to EDCA, only a single user receives data from the AP in a TXOP. In Fig.~\ref{fig_thr_EDCA} VO throughput increases proportionally to the $\alpha$ parameter and BE throughput decreases proportionally to $1-\alpha$.

\subsection{Two User Case}

\begin{figure*}[htb]
        \centering
        \begin{subfigure}[b]{0.45\textwidth}
                \includegraphics[width=\textwidth]{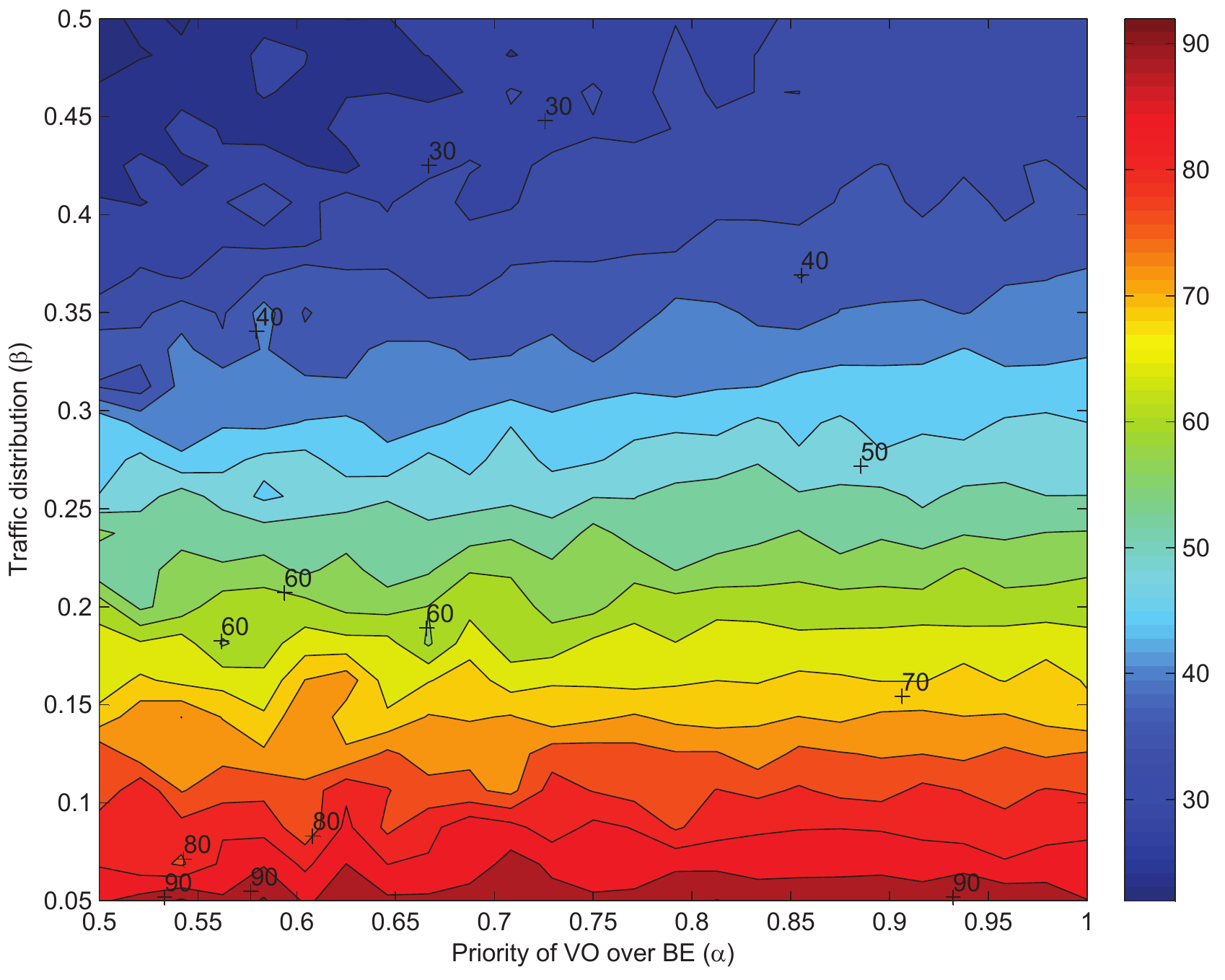}
             	\caption{\scriptsize{VO throughput change $T_{change}[VO]$ }}
                \label{VO_impr_2sta}
        \end{subfigure}%
\quad
        \begin{subfigure}[b]{0.45\textwidth}
                \includegraphics[width=\textwidth]{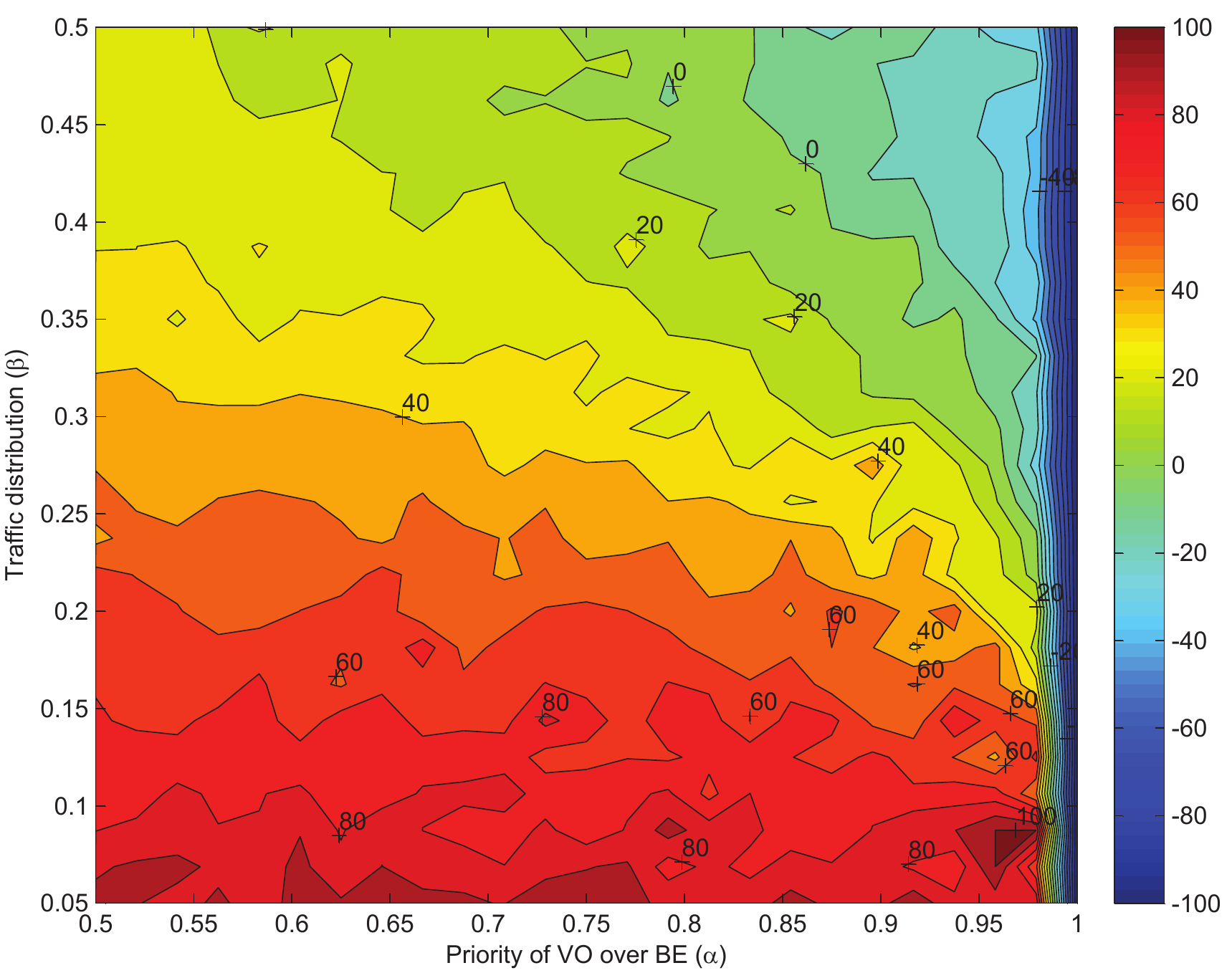}
		\caption{\scriptsize{BE throughput change  $T_{change}[BE]$}}
                 \label{BE_impr_2sta}
        \end{subfigure}
\label{thr_impr_2sta}
\caption{Throughput change comparison for traditional and DEMS queuing 
 as a function of $\alpha$ and  $\beta$ \big(cf. Eq.~(\ref{thr_ch})\big) for the two user case.
For $\alpha=1$ VO has strict priority over BE (cf. Ruckus SmartCast proposal \cite{ruckus}), for $\alpha=0.5$ there is no traffic differentiation, and for $\beta=0.5$ traffic destined to both users is uniformly distributed.
}
\end{figure*}

The results obtained for the two user case are presented in Fig.~\ref{VO_impr_2sta} and \ref{BE_impr_2sta} for  VO and  BE, respectively. The throughput change was calculated for varying values of $\alpha$ and $\beta$ separately for the two ACs according to Eq.~(\ref{thr_ch}). In the presented example, DEMS queuing always improves the average throughput of VO (the improvement ranges from $22\%$ to $92\%$, Fig.~\ref{VO_impr_2sta}). This also means that with DEMS the VO frames experience lower delays than with 802.11ac (i.e., they spend less time in queues). The improvement is the smallest when there are two users with the same probability of receiving frames ($\beta=0.5$) and the probability of selecting frames from BE and VO queues is similar ($\alpha=0.5$), i.e., without traffic differentiation (with a standard-compliant EDCA configuration this is purely a theoretical case). For $\alpha \leq 0.77$  the throughput of BE is also improved by DEMS (up to $100\%$, Fig.~\ref{BE_impr_2sta}). For other $\alpha$ values, under small $\beta$, BE throughput can also be improved. Otherwise it is worsened at the expense of VO throughput improvement.

\begin{figure*}[htb]
        \centering
        \begin{subfigure}[b]{0.45\textwidth}
                \includegraphics[width=\textwidth]{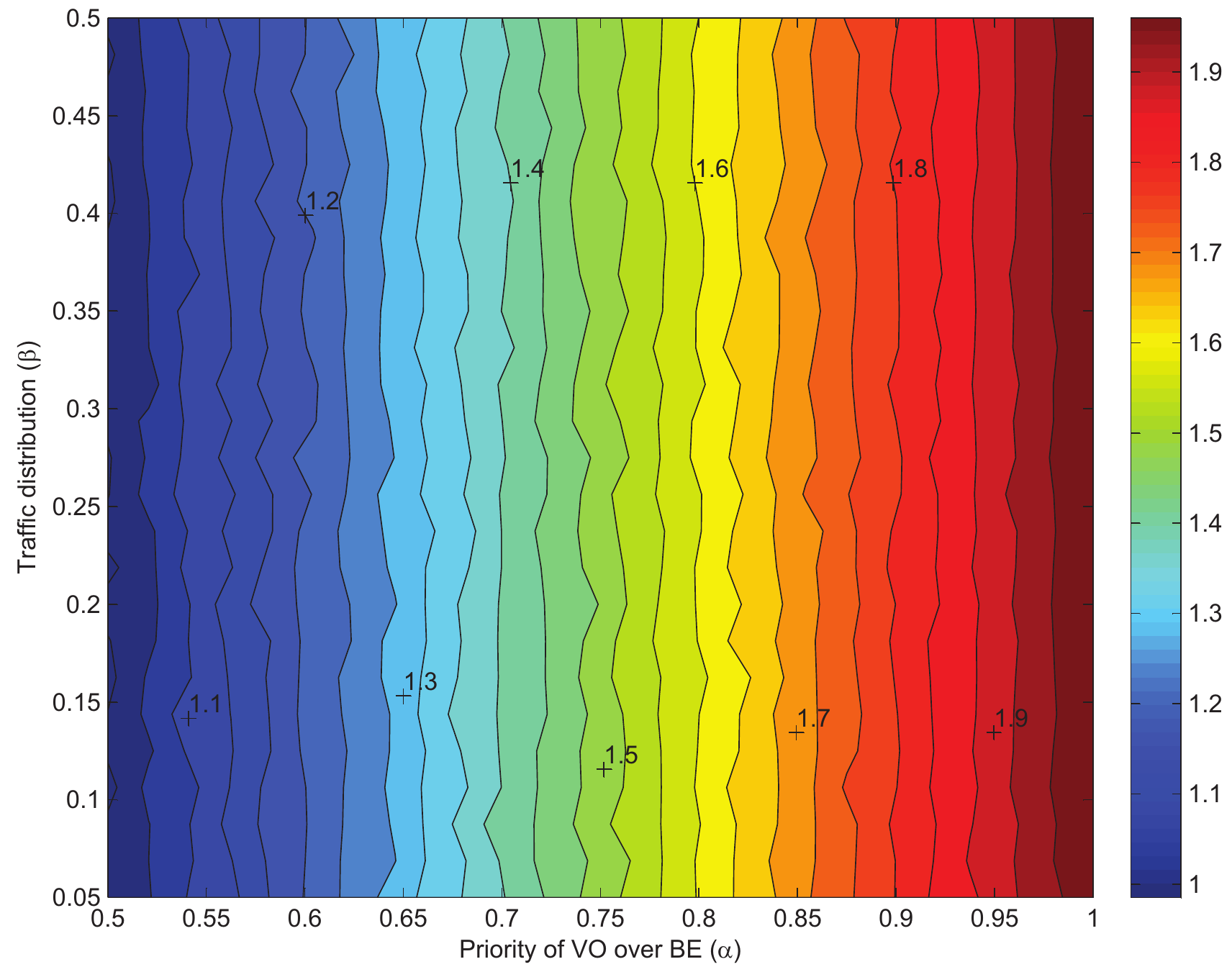}
                \caption{\scriptsize{VO throughput \big($c_{\text{DEMS}}[VO](\alpha,\beta)$\big) for DEMS}}
                \label{VO_DEMS_2sta}
        \end{subfigure}%
        \quad 
        \begin{subfigure}[b]{0.45\textwidth}
                \includegraphics[width=\textwidth]{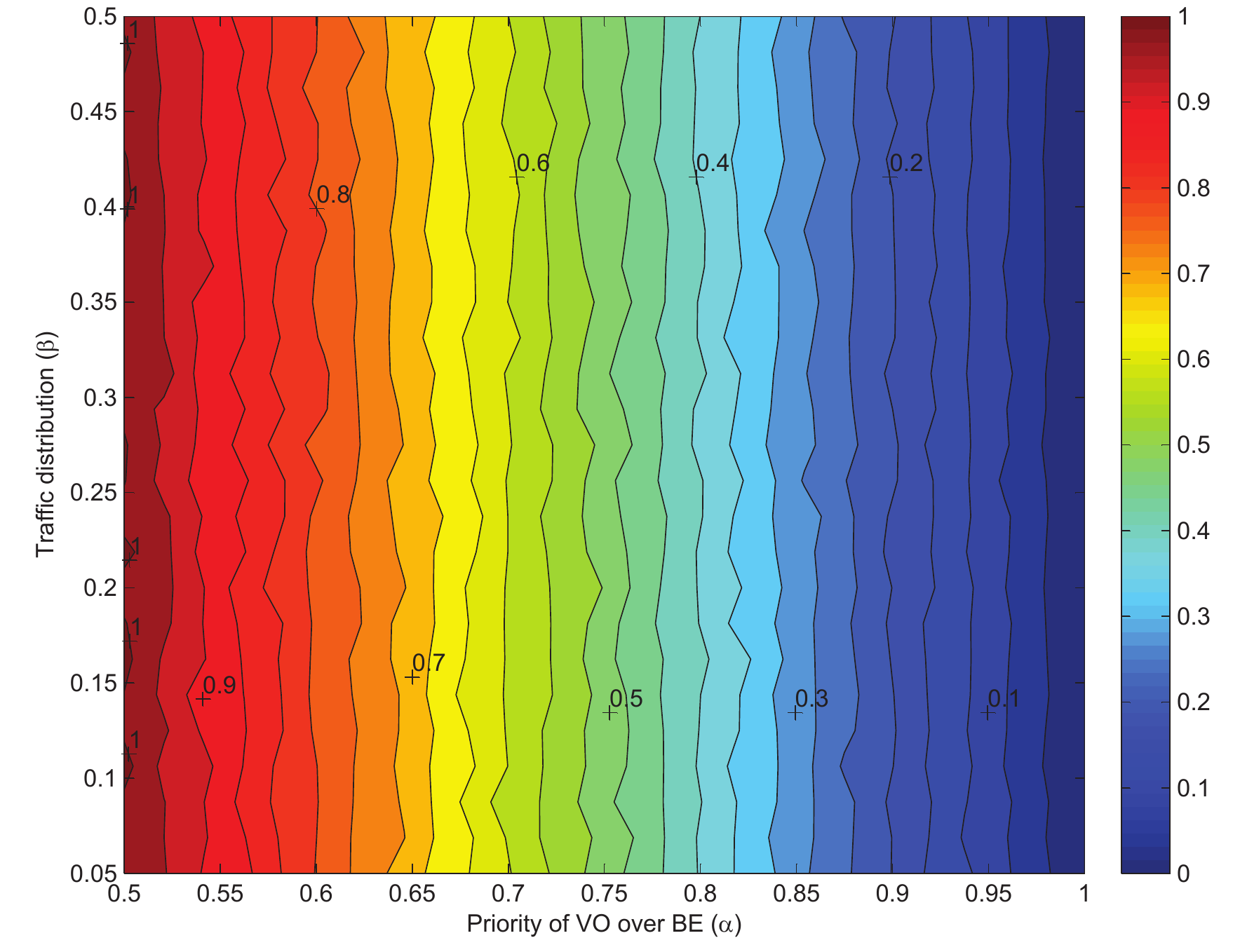}
		\caption{\scriptsize{BE throughput \big($c_{\text{DEMS}}[BE](\alpha,\beta)$\big) for DEMS}}
                 \label{BE_DEMS_2sta}
        \end{subfigure}
\break 
 \begin{subfigure}[b]{0.45\textwidth}
               \includegraphics[width=\textwidth]{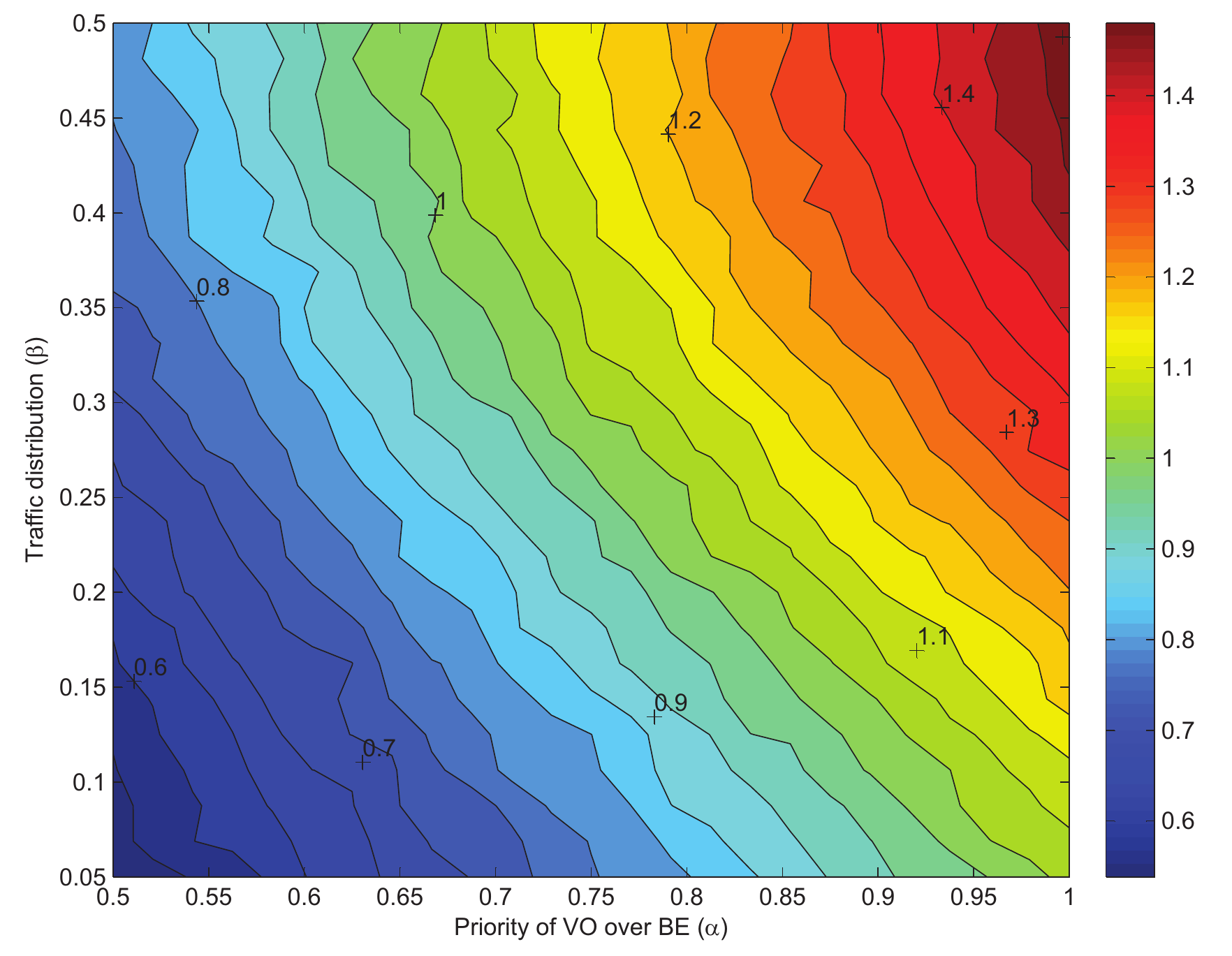}
                \caption{\scriptsize{VO throughput \big($c_{\text{802.11ac}}[VO](\alpha,\beta)$\big) for 802.11ac}}
                \label{VO_80211ac_2sta}
        \end{subfigure}%
        \quad 
        \begin{subfigure}[b]{0.45\textwidth}
               \includegraphics[width=\textwidth]{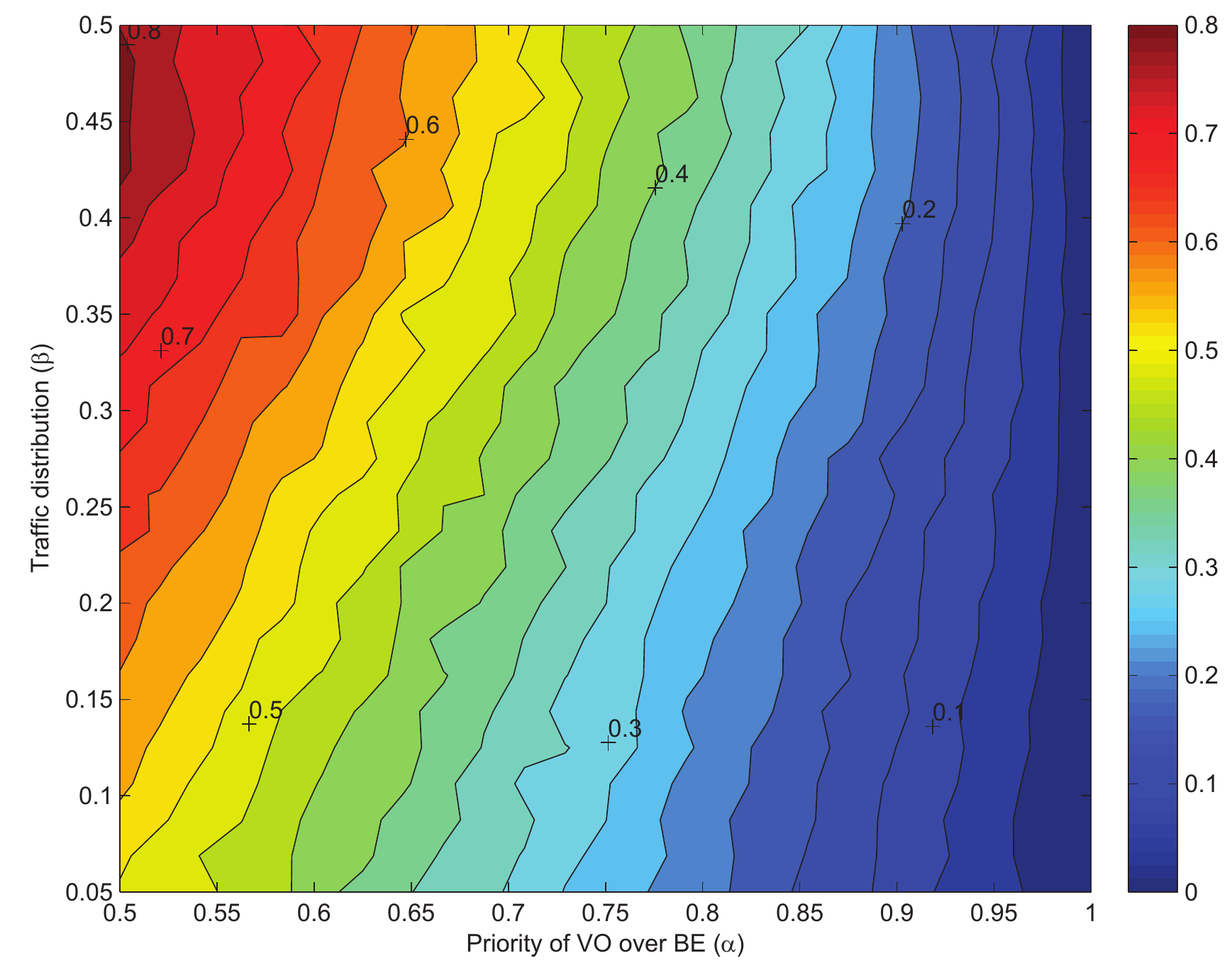}
		\caption{\scriptsize{BE throughput \big($c_{\text{802.11ac}}[BE](\alpha,\beta)$\big) for 802.11ac}}
                 \label{BE_80211ac_2sta}
        \end{subfigure}
\caption{Comparison of the normalized throughput for the two user case as a function of $\alpha$ and $\beta$.
$c[i](\alpha,\beta)=2$ means that two frames of the $i$-th AC are transmitted simultaneously in the shared TXOP and they are destined to two non-interfering receivers. 
}
\label{thr_2sta}
\end{figure*}

Fig.~\ref{thr_2sta}  shows the normalized  VO and BE throughput for DEMS \big($c_{\text{DEMS}}[VO](\alpha,\beta)$ and $c_{\text{DEMS}}[BE](\alpha,\beta)$, respectively\big) and for traditional scheduling \big($c_{\text{802.11ac}}[VO](\alpha,\beta)$ and $c_{\text{802.11ac}}[BE](\alpha,\beta)$, respectively\big).
For DEMS, the larger the possibility of selecting the VO frame by the scheduler $\alpha$ the higher the normalized throughput of VO and the lower the normalized throughput of BE. For 802.11ac the normalized throughput of both ACs is also highly dependent on the traffic distribution $\beta$, the more uniform  the traffic distribution the higher the throughput. 

\begin{figure*}[htb]
        \centering
        \begin{subfigure}[b]{0.4\textwidth}
                \includegraphics[width=\textwidth]{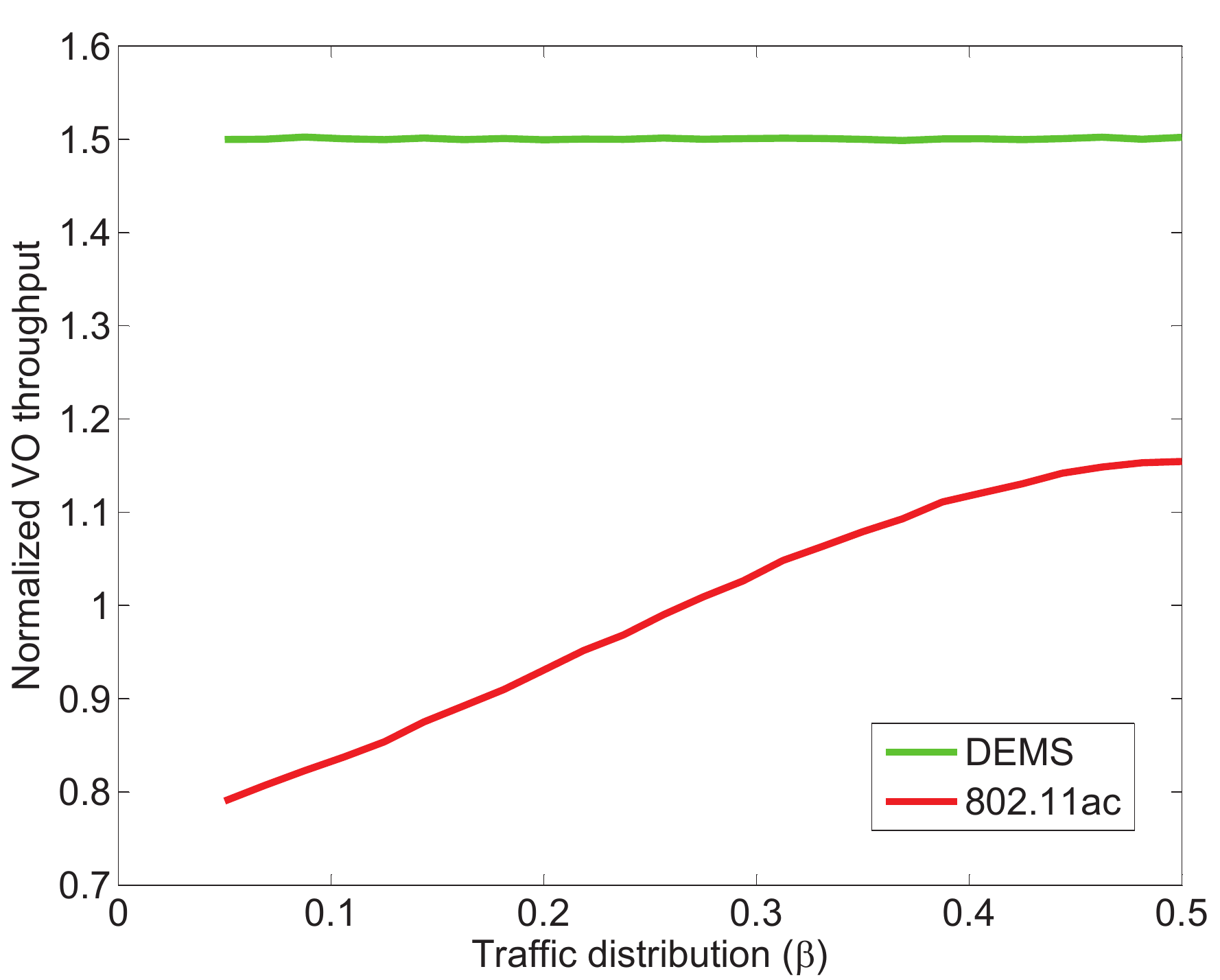}
                \caption{\scriptsize{Average VO throughput $T_{avg}[VO](\beta)$}}
\quad
                \label{VO_comp_2sta}
        \end{subfigure}
        \quad 
        \begin{subfigure}[b]{0.4\textwidth}
                \includegraphics[width=\textwidth]{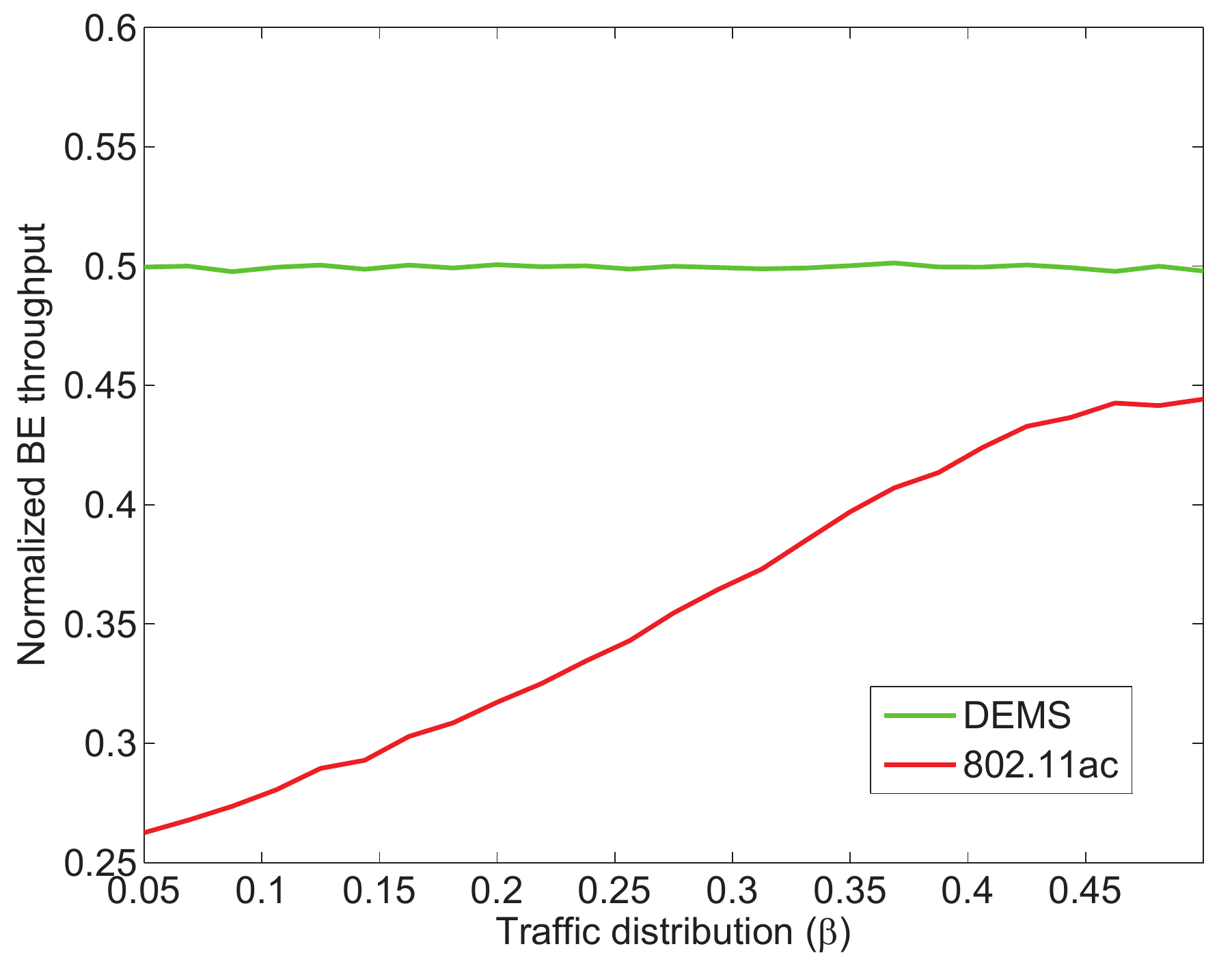}
		\caption{\scriptsize{Average BE throughput $T_{avg}[BE](\beta)$}}
\quad
                 \label{BE_comp_2sta}
        \end{subfigure}
        \quad
        \begin{subfigure}[b]{0.4\textwidth}
                \includegraphics[width=\textwidth]{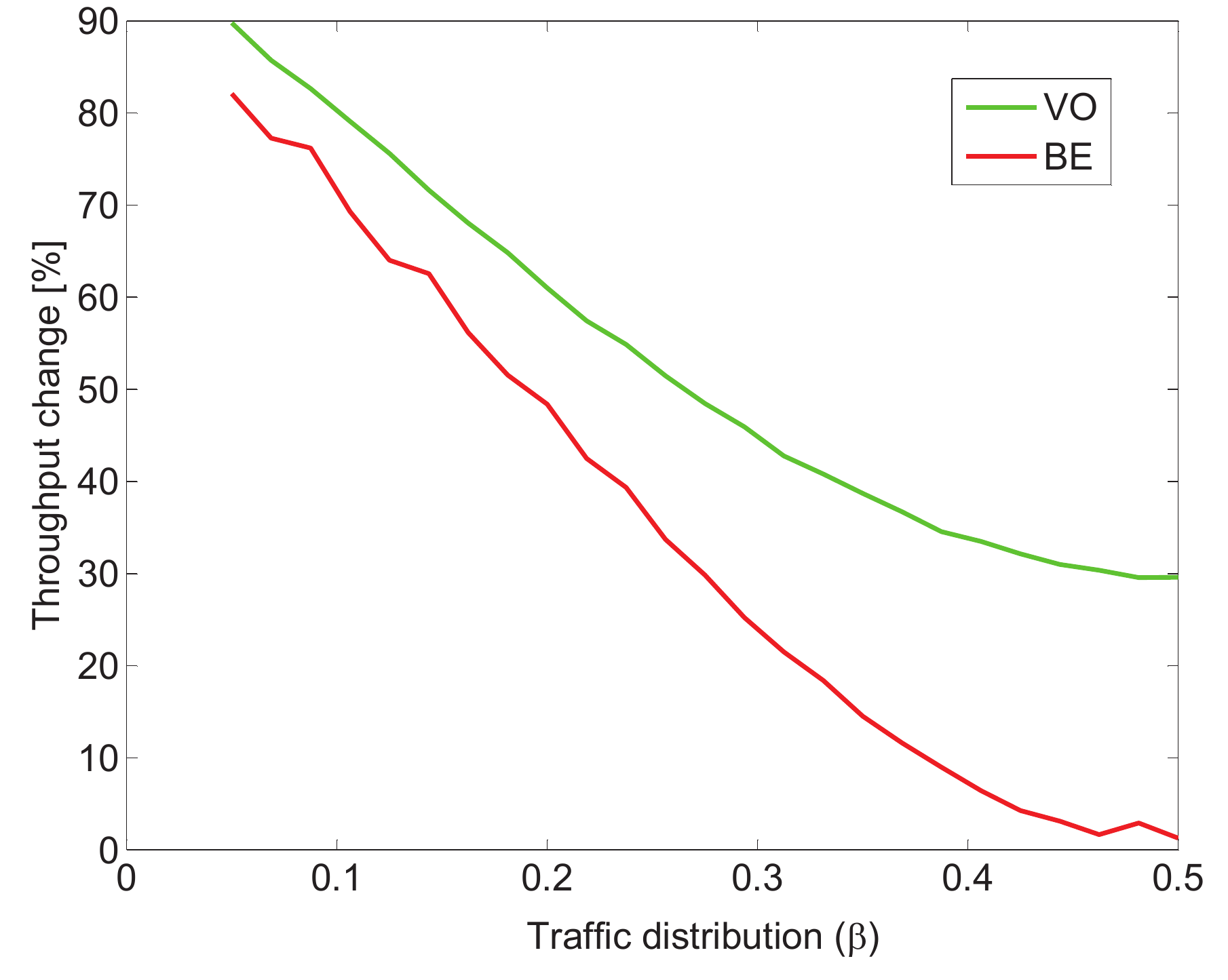}
		\caption{\scriptsize{DEMS average throughput change $T_{change}^{avg}[i](\beta)$ in comparison to 802.11ac}}
                 \label{VOBE_comp_2sta}
        \end{subfigure}
\caption{Comparison of the average throughput for DEMS and traditional scheduling as a function of $\beta$ for the two user case.
}
\label{comp_2sta}
\end{figure*}

\newpage
The comparison of the average VO and BE normalized throughput values $T_{avg}[i]$ as a function of $\beta$ 
for DEMS and 802.11ac is shown in Fig.~\ref{VO_comp_2sta} and Fig.~\ref{BE_comp_2sta}, respectively. For DEMS the average VO and BE throughput value is stable for every $\beta$ and equals 1.5 and 0.5, respectively. This means that on average 1.5 VO and 0.5 BE frames were transmitted in a single MU-MIMO transmission period. For 802.11ac, the more similar the probability of receiving frames by the two users, the higher the average VO and BE throughput values because non-uniform traffic distribution increases the HOL blocking probability. The minimum VO (BE) average throughput is 0.79 (0.26) and the maximum VO (BE) average throughput is 1.15 (0.44). 
Fig.~\ref{VOBE_comp_2sta}  additionally presents the average VO and BE DEMS throughput change $T_{change}^{avg}$ as a function of $\beta$ in comparison to 802.11ac. The average throughput change is positive: for VO it ranges from $30\%$ to $90\%$ and for BE it ranges from $2\%$ to $82\%$. 

\begin{figure*}[htb]
        \centering
        \begin{subfigure}[b]{0.4\textwidth}
                \includegraphics[width=\textwidth]{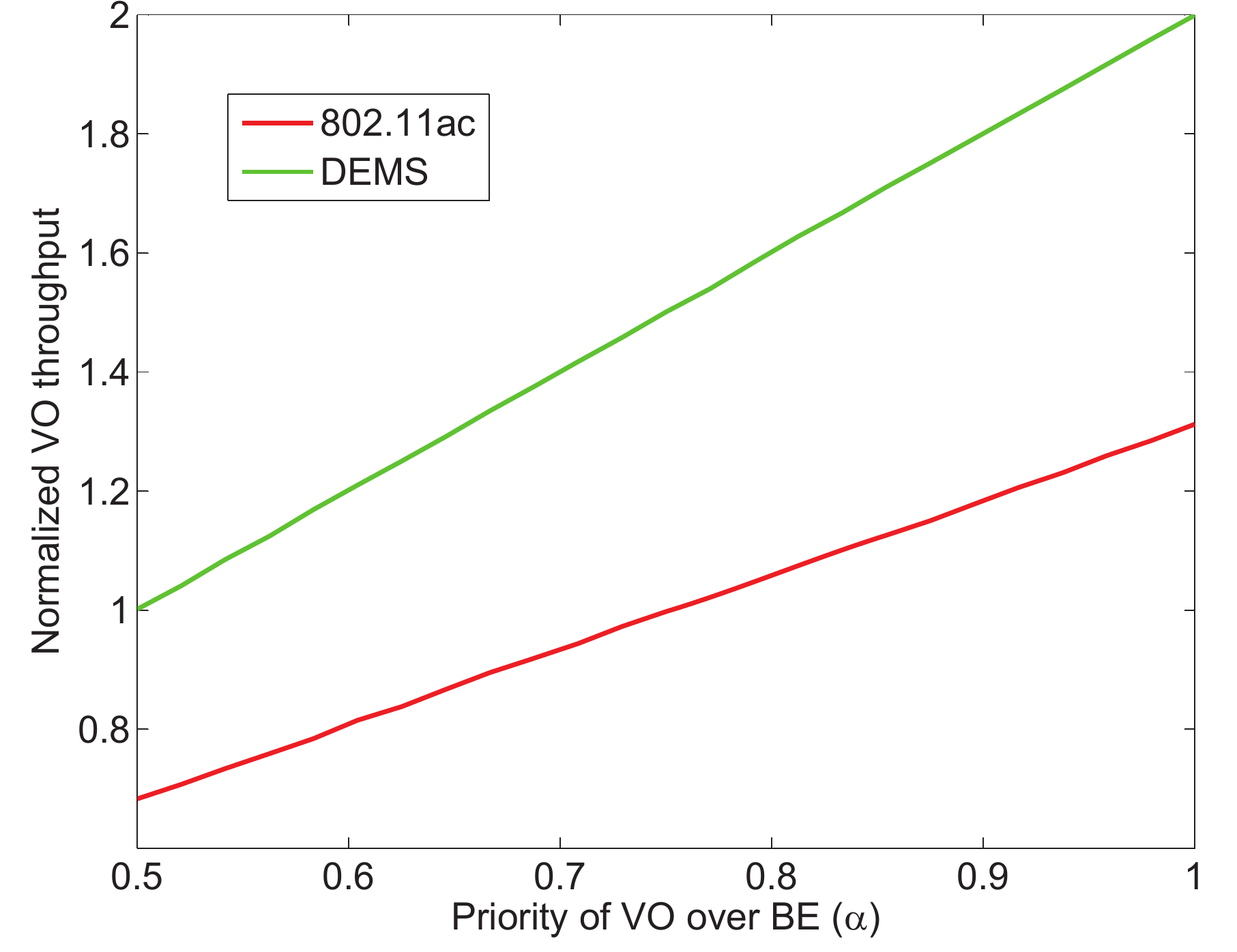}
                \caption{\scriptsize{Average VO throughput $T_{avg}[VO](\alpha)$}}
\quad
                \label{VO_comp_2sta-alpha}
        \end{subfigure}%
        \quad 
        \begin{subfigure}[b]{0.4\textwidth}
                \includegraphics[width=\textwidth]{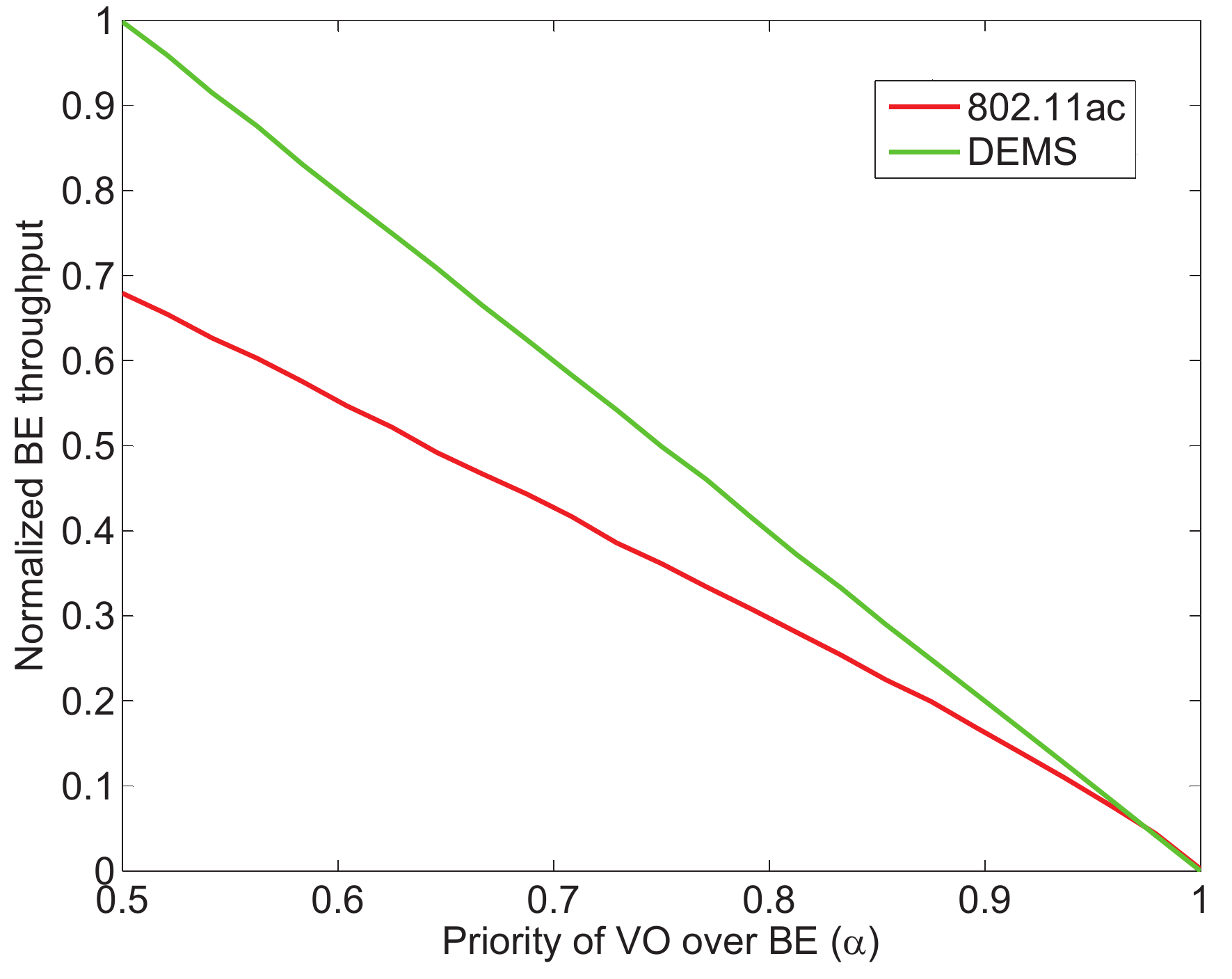}
		\caption{\scriptsize{Average BE throughput $T_{avg}[BE](\alpha)$}}
\quad
                 \label{BE_comp_2sta-alpha}
        \end{subfigure}
        \quad
        \begin{subfigure}[b]{0.4\textwidth}
                \includegraphics[width=\textwidth]{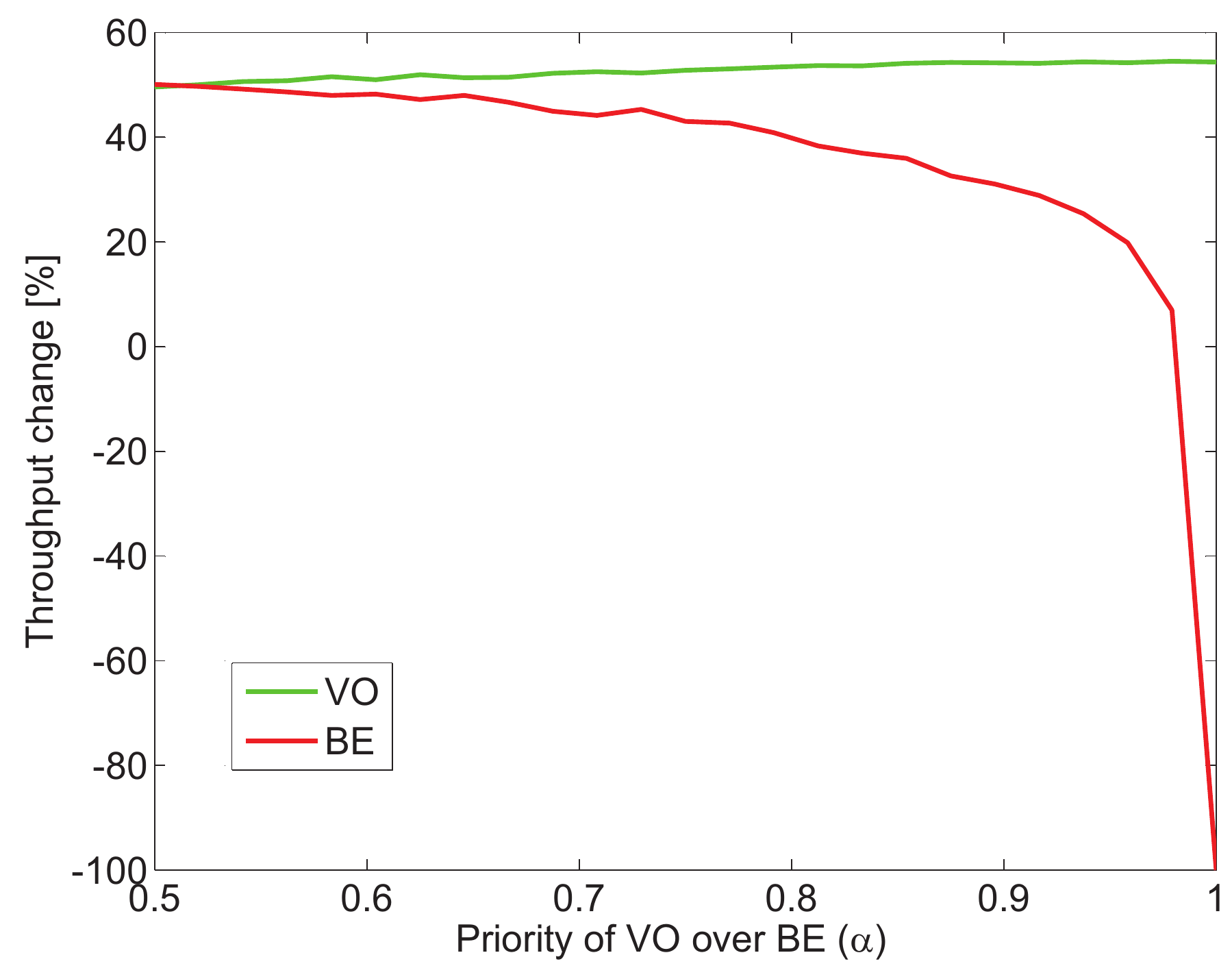}
		\caption{\scriptsize{DEMS average throughput change $T_{change}^{avg}[i](\alpha)$ in comparison to 802.11ac}}
                 \label{VOBE_comp_2sta-alpha}
        \end{subfigure}
\caption{Comparison of the average throughput for DEMS and traditional scheduling as a function of $\alpha$ for the two user case.
}
\label{comp_2sta-alpha}
\end{figure*}

The comparison of the average VO and BE normalized throughput $T_{avg}[i]$ as a function of $\alpha$ for DEMS and 802.11ac is shown in Fig.~\ref{VO_comp_2sta-alpha} and Fig.~\ref{BE_comp_2sta-alpha}, respectively. For DEMS the average VO throughput increases from 1 to 2 and BE throughput decreases from 1 to 0 for increasing $\alpha$. For 802.11ac the average VO throughput increases from 0.68 to 1.31 and BE throughput values decrease from 0.68 to 0.002 for increasing $\alpha$, which means that with $\alpha=0.5$ (no traffic differentiation) $32\%$ and for $\alpha=1$ (strict VO priority over BE) $34\%$ of simultaneous transmission opportunities are wasted due to HOL blocking. 
Fig.~\ref{VOBE_comp_2sta-alpha}  additionally presents the average VO and BE DEMS throughput values $T_{change}^{avg}$ as a function of $\alpha$ 
in comparison to 802.11ac. The average throughput change is positive and almost stable for VO (ranging from $49.6\%$ to $54.5\%$). For BE it decreases from $50.1\%$ to $-100\%$ for the increasing $\alpha$.

\subsection{Three User Case}

\begin{figure*}[htb]
        \centering
        \begin{subfigure}[b]{0.45 \textwidth}
                \includegraphics[width=\textwidth]{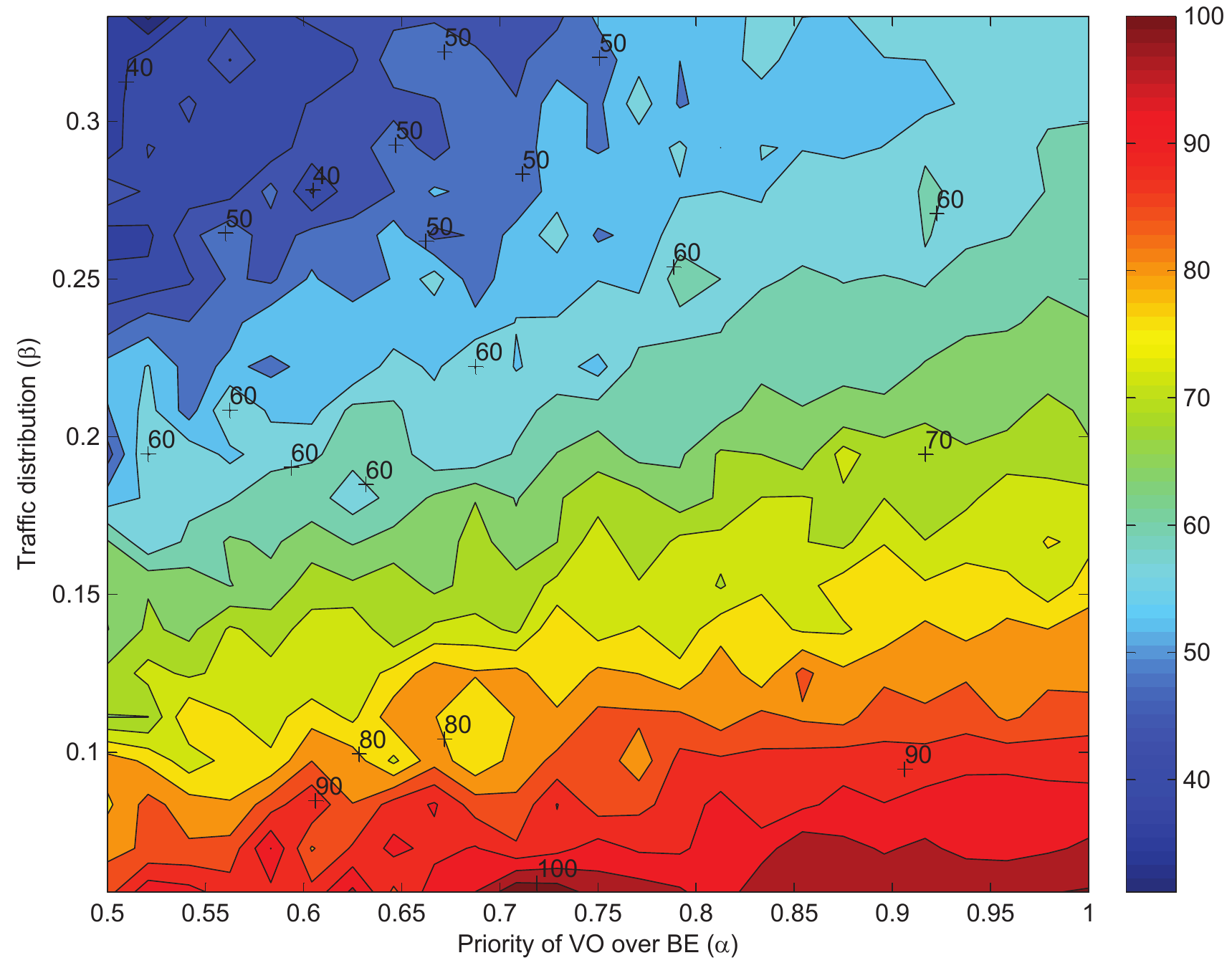}
                \caption{\scriptsize{VO throughput change $T_{change}[VO]$}}
                \label{VO_impr_3sta}
        \end{subfigure}%
\quad
        \begin{subfigure}[b]{0.45 \textwidth}
                \includegraphics[width=\textwidth]{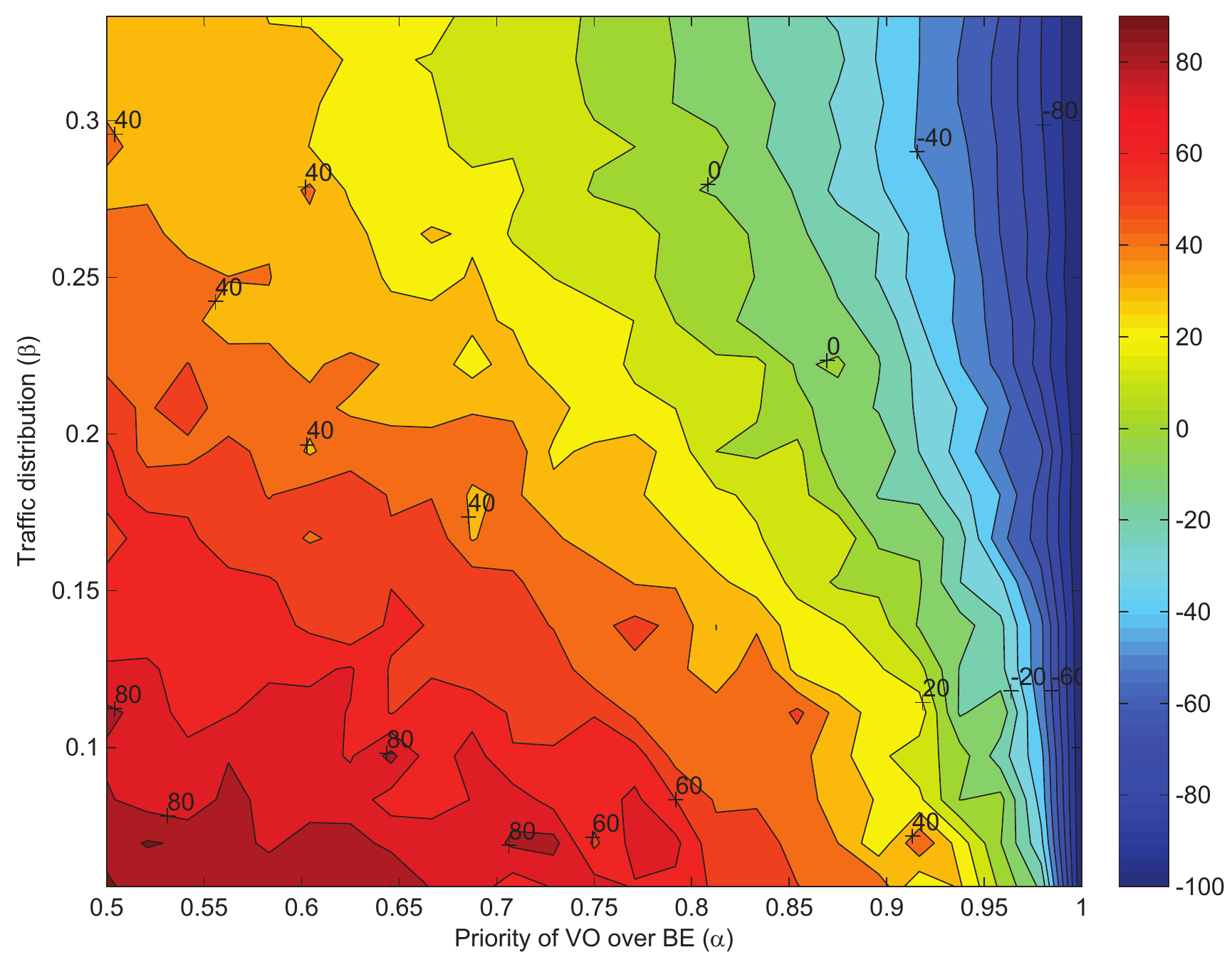}
		\caption{\scriptsize{BE throughput change $T_{change}[BE]$}}
                 \label{BE_impr_3sta}
        \end{subfigure}
\caption{Throughput change comparison for traditional and DEMS queuing as a function of $\alpha$ and  $\beta$ \big(cf. Eq.~(\ref{thr_ch})\big) for the three user case.
For $\alpha=1$ VO frames have strict priority over BE frames (cf. Ruckus SmartCast proposal \cite{ruckus}), for $\alpha=0.5$ there is no traffic differentiation, and for $\beta=1/3$ traffic destined to all three users is uniformly distributed.
}
\end{figure*}

The  results obtained for the three user case are presented in Fig.~\ref{VO_impr_3sta} and \ref{BE_impr_3sta} for VO and  BE, respectively. The throughput change was calculated for varying values of $\alpha$ and $\beta$ separately for the two ACs according to Eq.~(\ref{thr_ch}). The observed behavior confirms the results from the two user case. In the presented example, DEMS queuing always improves the average throughput of VO (by $31\%$ to $100\%$, cf. Fig.~\ref{VO_impr_3sta}). Therefore, in a given time period from $31\%$ to $100\%$ more frames are transmitted by DEMS than by 802.11ac, which also means that the average VO queuing delay decreases, which is important for delay-sensitive traffic. The improvement is the smallest when there are three users with the same probability of receiving frames ($\beta=1/3$) and the probability of selecting frames from BE and VO queues is similar (for  $\alpha=0.5$), i.e., without traffic differentiation. This can be explained by the fact that the more uniform traffic distribution the lower probability of HOL blocking, therefore, the difference between DEMS and 802.11ac is the smallest. For $\alpha \leq 0.77$  the throughput of BE is also improved by DEMS (up to $90\%$, cf. Fig.~\ref{BE_impr_3sta}). For other values of $\alpha$, but under small $\beta$, BE throughput can be also improved. Otherwise, it is worsened at the expense of VO throughput improvement. 

\begin{figure*}[htb]
        \centering
        \begin{subfigure}[b]{0.45\textwidth}
                \includegraphics[width=\textwidth]{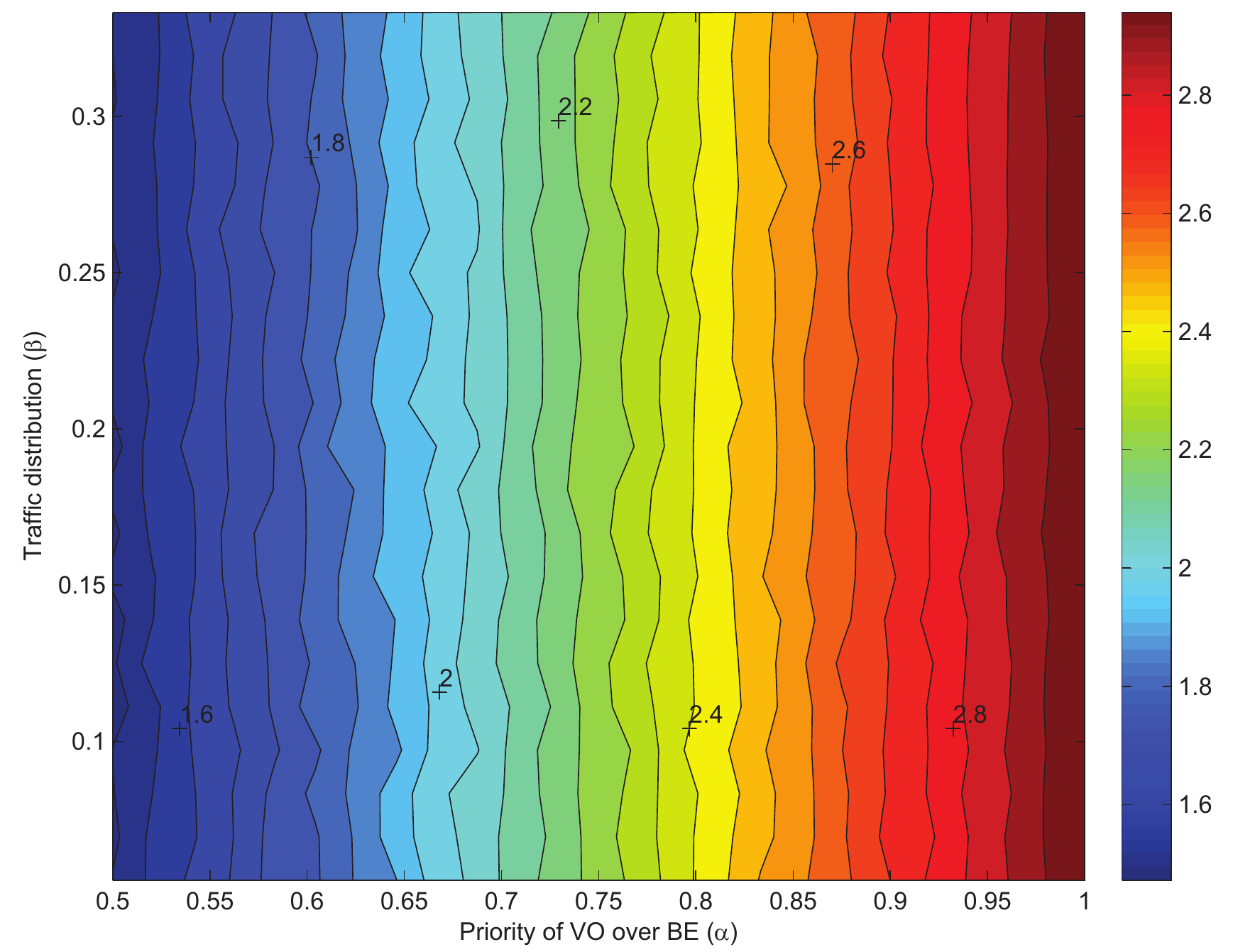}
                \caption{\scriptsize{VO throughput \big($c_{\text{DEMS}}[VO](\alpha,\beta)$\big) for DEMS}}
                \label{VO_DEMS_3sta}
        \end{subfigure}%
        \quad
        \begin{subfigure}[b]{0.45\textwidth}
             \includegraphics[width=\textwidth]{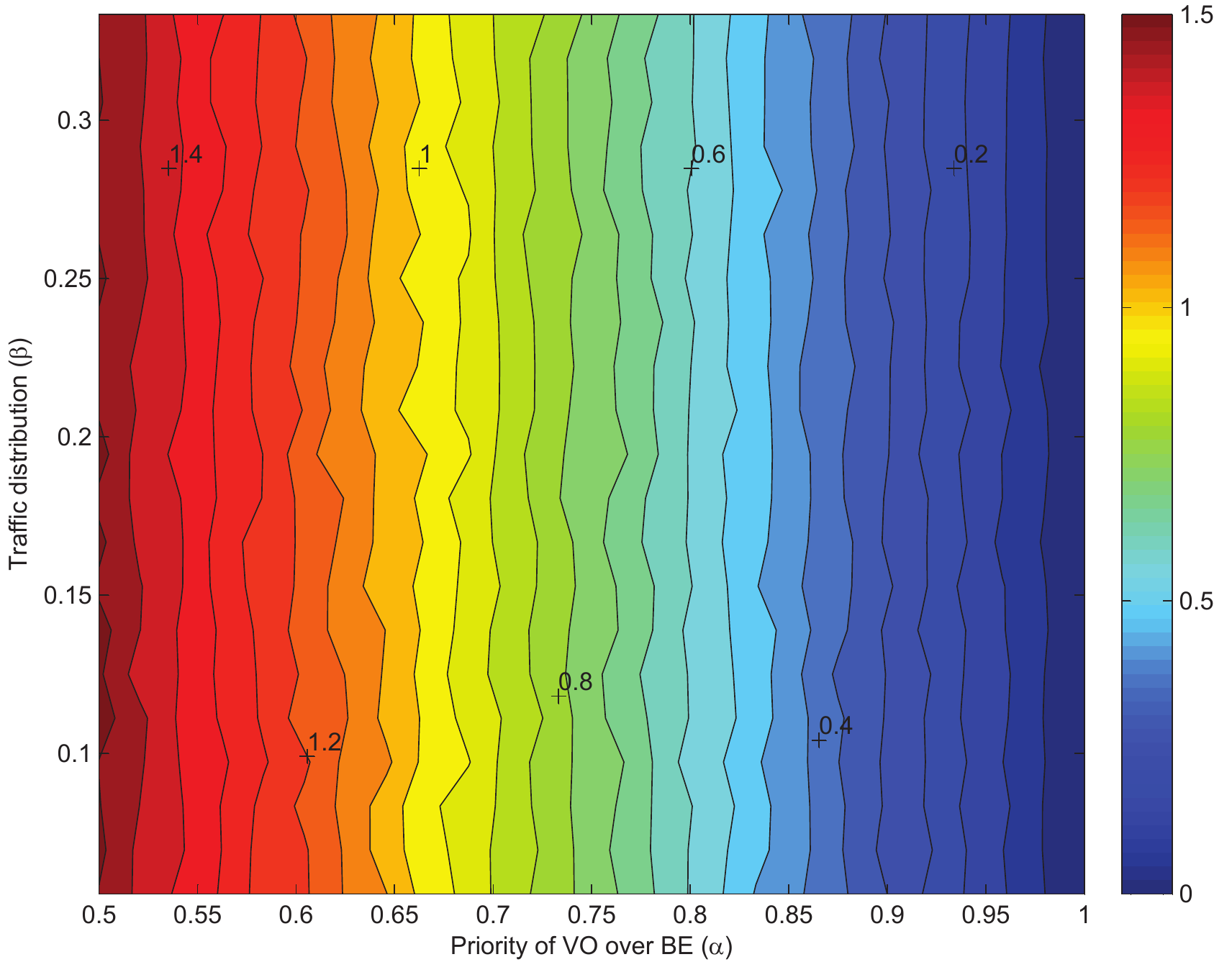}
		\caption{\scriptsize{BE throughput \big($c_{\text{DEMS}}[BE](\alpha,\beta)$\big) for DEMS}}
                 \label{BE_DEMS_3sta}
        \end{subfigure}
\break
 \begin{subfigure}[b]{0.45\textwidth}
          \includegraphics[width=\textwidth]{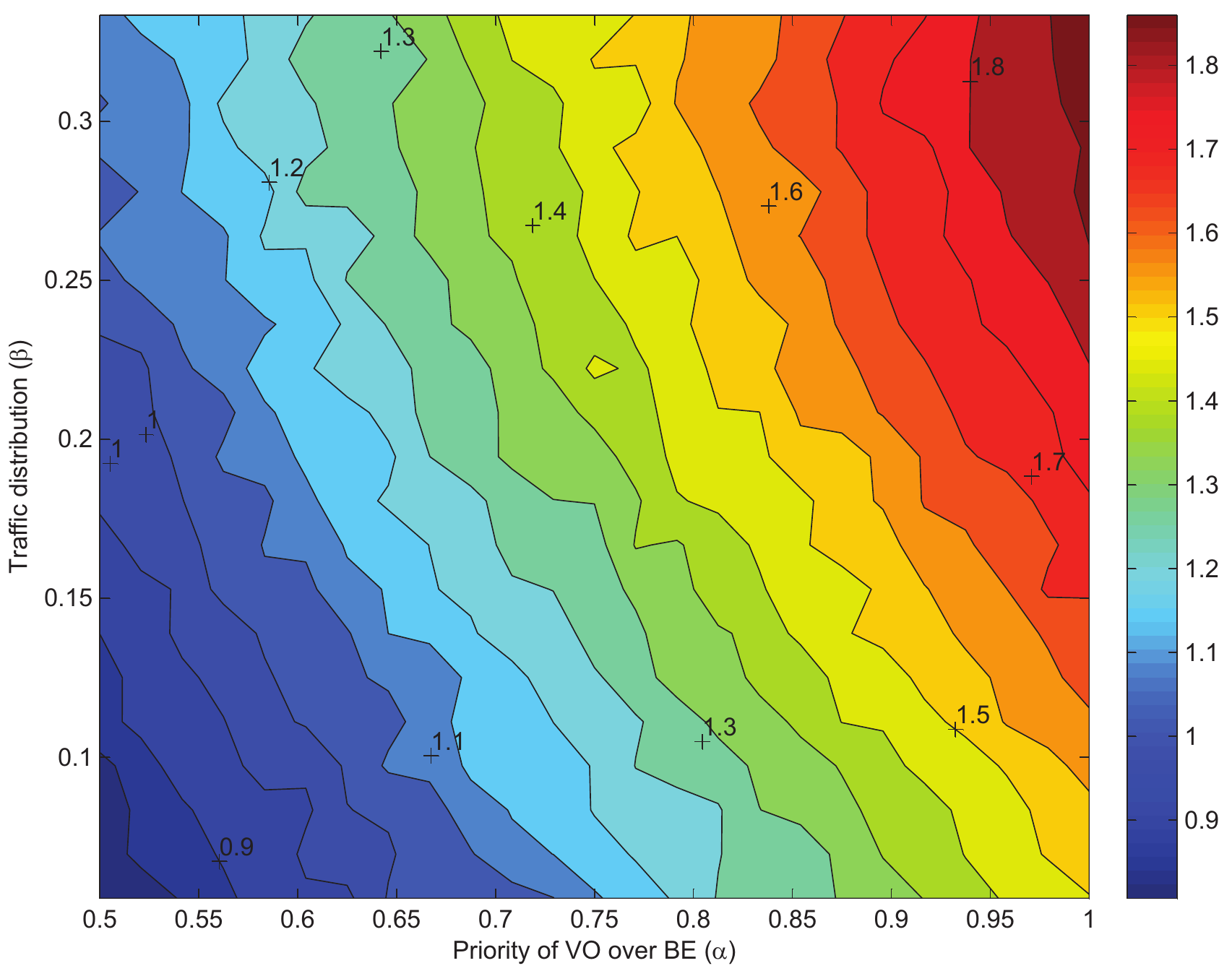}
                \caption{\scriptsize{VO throughput \big($c_{\text{802.11ac}}[VO](\alpha,\beta)$\big) for 802.11ac}}
                \label{VO_80211ac_3sta}
        \end{subfigure}%
        \quad 
        \begin{subfigure}[b]{0.45\textwidth}
       \includegraphics[width=\textwidth]{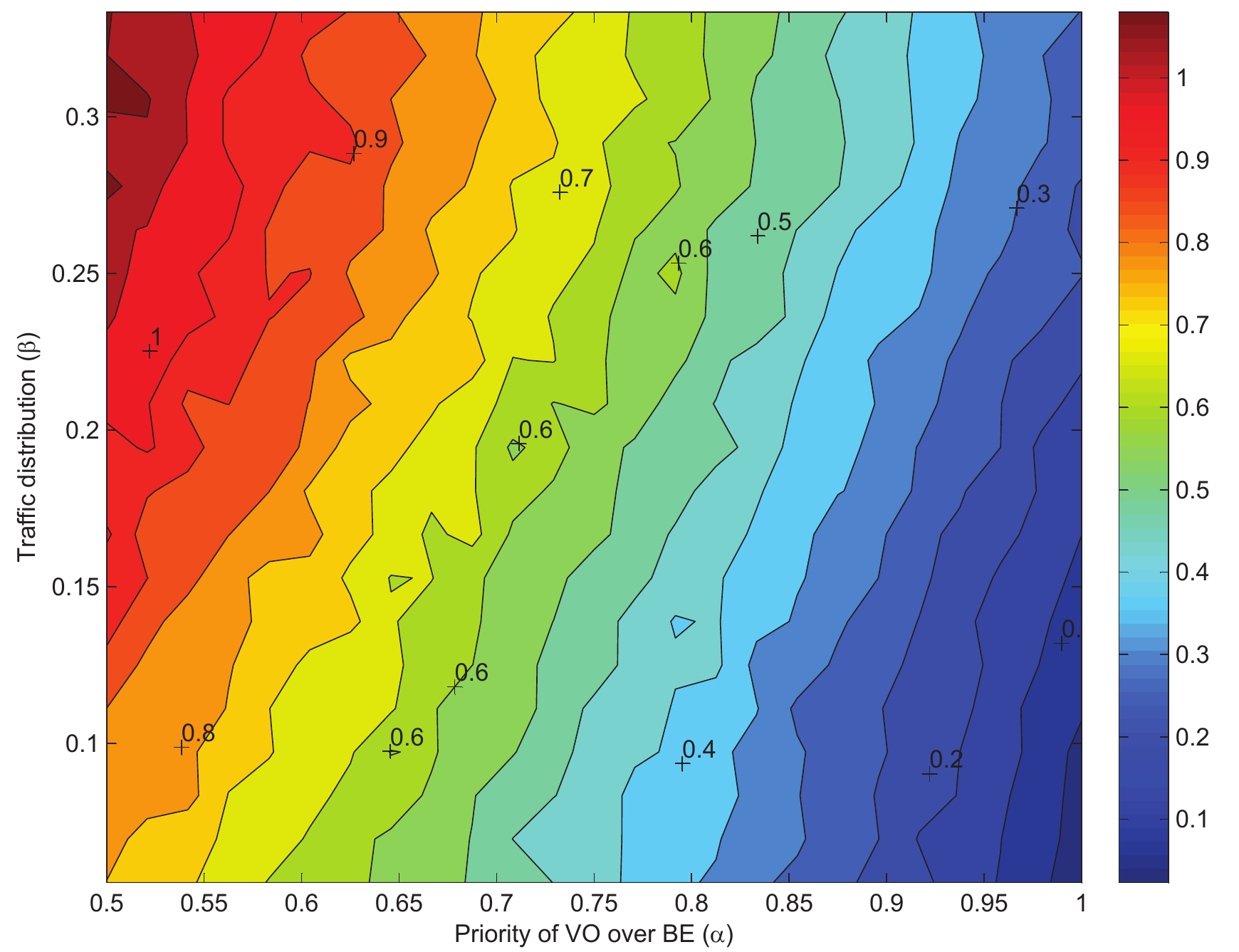}
		\caption{\scriptsize{BE throughput \big($c_{\text{802.11ac}}[BE](\alpha,\beta)$\big) for 802.11ac}}
                 \label{BE_802.11ac_3sta}
        \end{subfigure}
\caption{Comparison of the normalized throughput for the three user case as a function of $\alpha$ and $\beta$. $c[i](\alpha,\beta)=3$ means that three frames of the $i$-th AC are transmitted simultaneously in the shared TXOP and they are destined to two non-interfering receivers. 
}
\label{thr_3sta}
\end{figure*}

Fig.~\ref{thr_3sta} additionally shows the VO (BE) normalized throughput for an AP with DEMS queuing $c_{\text{DEMS}}[VO](\alpha,\beta)$ \big($c_{\text{DEMS}}[BE](\alpha,\beta)$\big) and for an AP with traditional scheduling $c_{\text{802.11ac}}[VO](\alpha,\beta)$ \big($c_{\text{802.11ac}}[BE](\alpha,\beta)$\big). The observed behavior confirms the results from the two user case. For DEMS, the larger the possibility of selecting the VO frame by the  scheduler $\alpha$ the higher the normalized throughput of VO and the lower the normalized throughput of BE. For 802.11ac the normalized throughput of both ACs is additionally highly dependent on the traffic distribution $\beta$, the more uniform is the traffic distribution among the MUs the higher the throughput values since the HOL blocking probability is the smallest. 

\begin{figure*}[htb]
        \centering
        \begin{subfigure}[b]{0.4 \textwidth}
                \includegraphics[width=\textwidth]{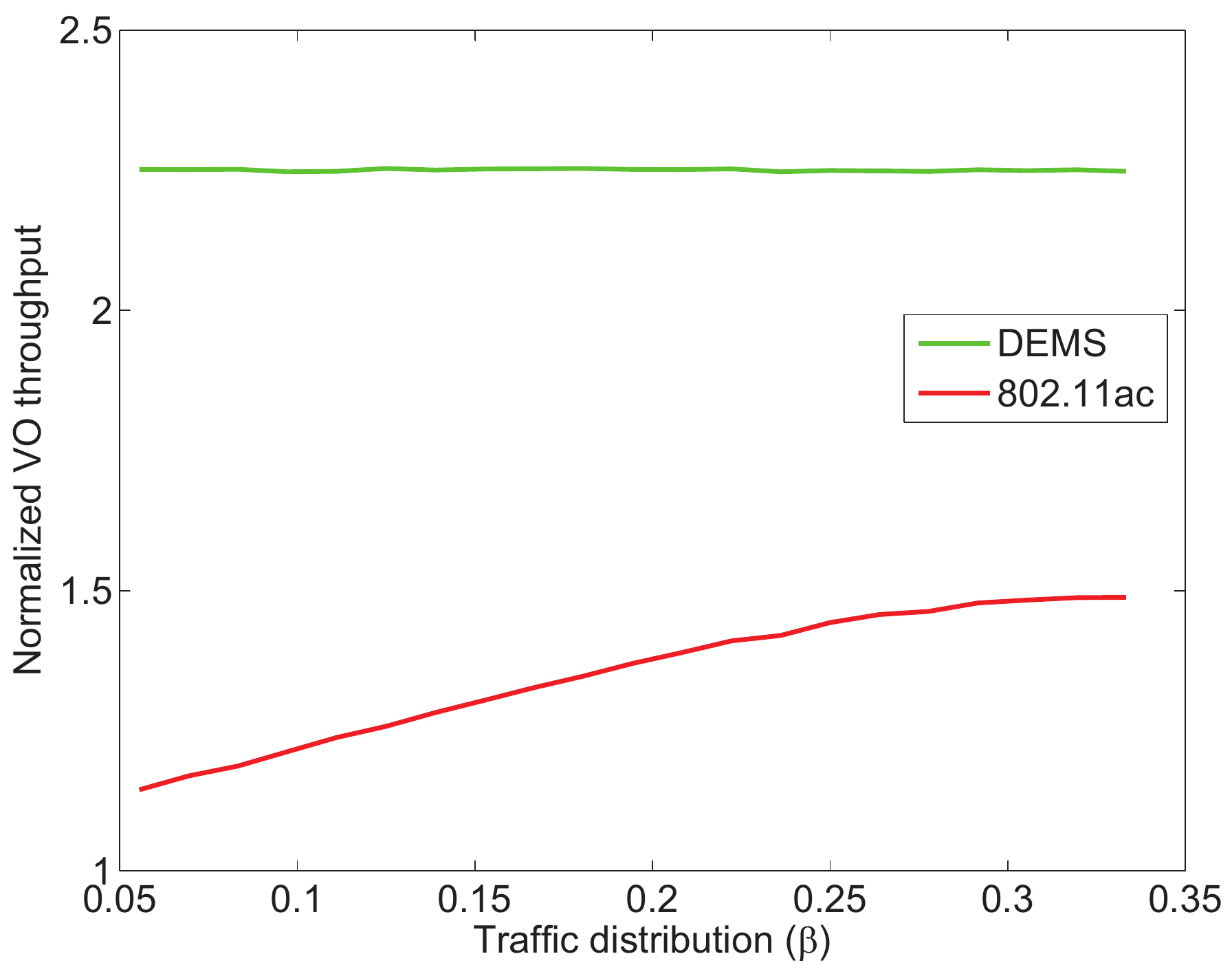}
                \caption{\scriptsize{Average VO throughput $T_{avg}[VO](\beta)$}}
        \quad 
                \label{VO_comp_3sta}
        \end{subfigure}%
        \quad 
        \begin{subfigure}[b]{0.4 \textwidth}
                \includegraphics[width=\textwidth]{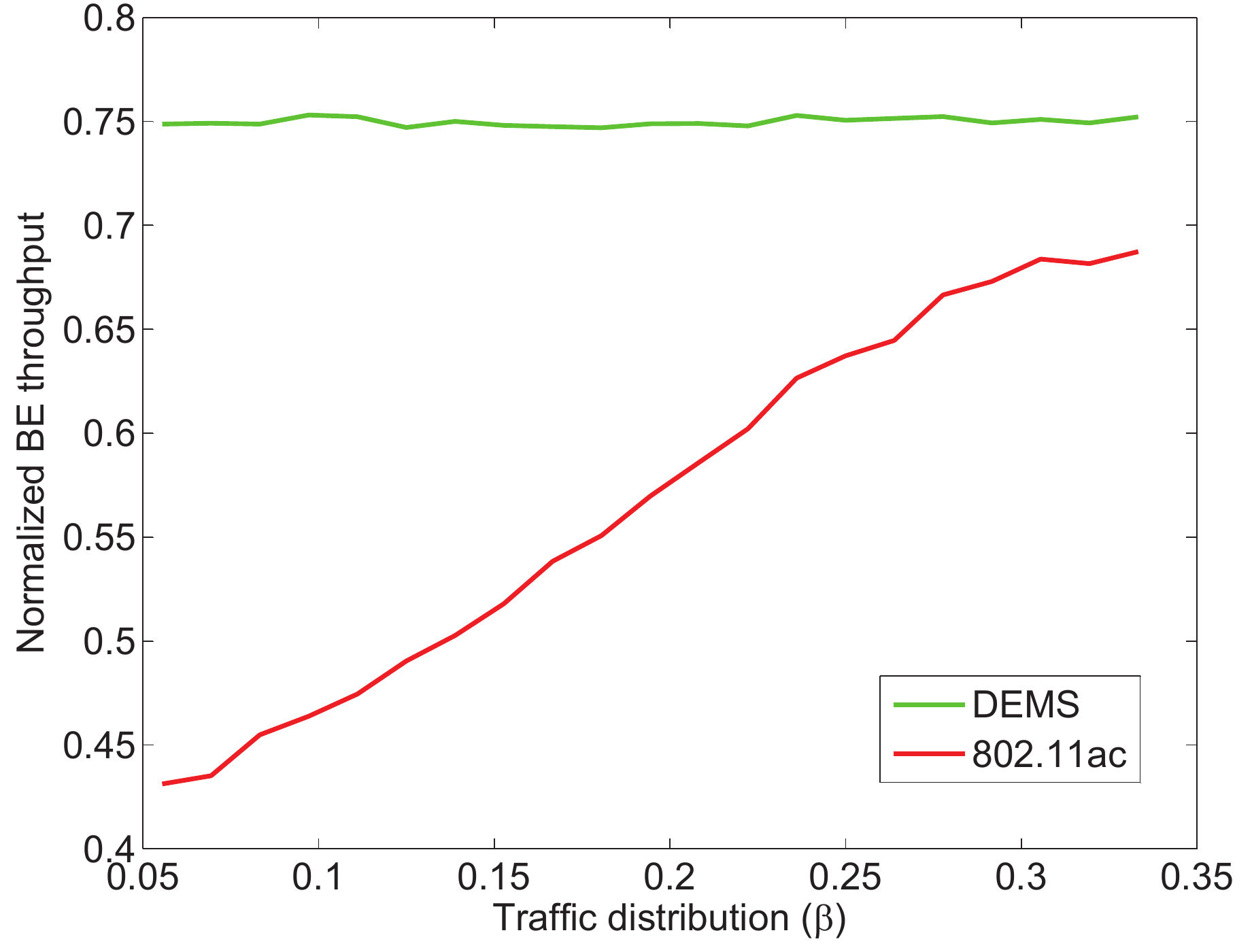}
		\caption{\scriptsize{Average BE throughput $T_{avg}[BE](\beta)$}}
        \quad 
                 \label{BE_comp_3sta}
        \end{subfigure}
        \quad 
        \begin{subfigure}[b]{0.4 \textwidth}
                \includegraphics[width=\textwidth]{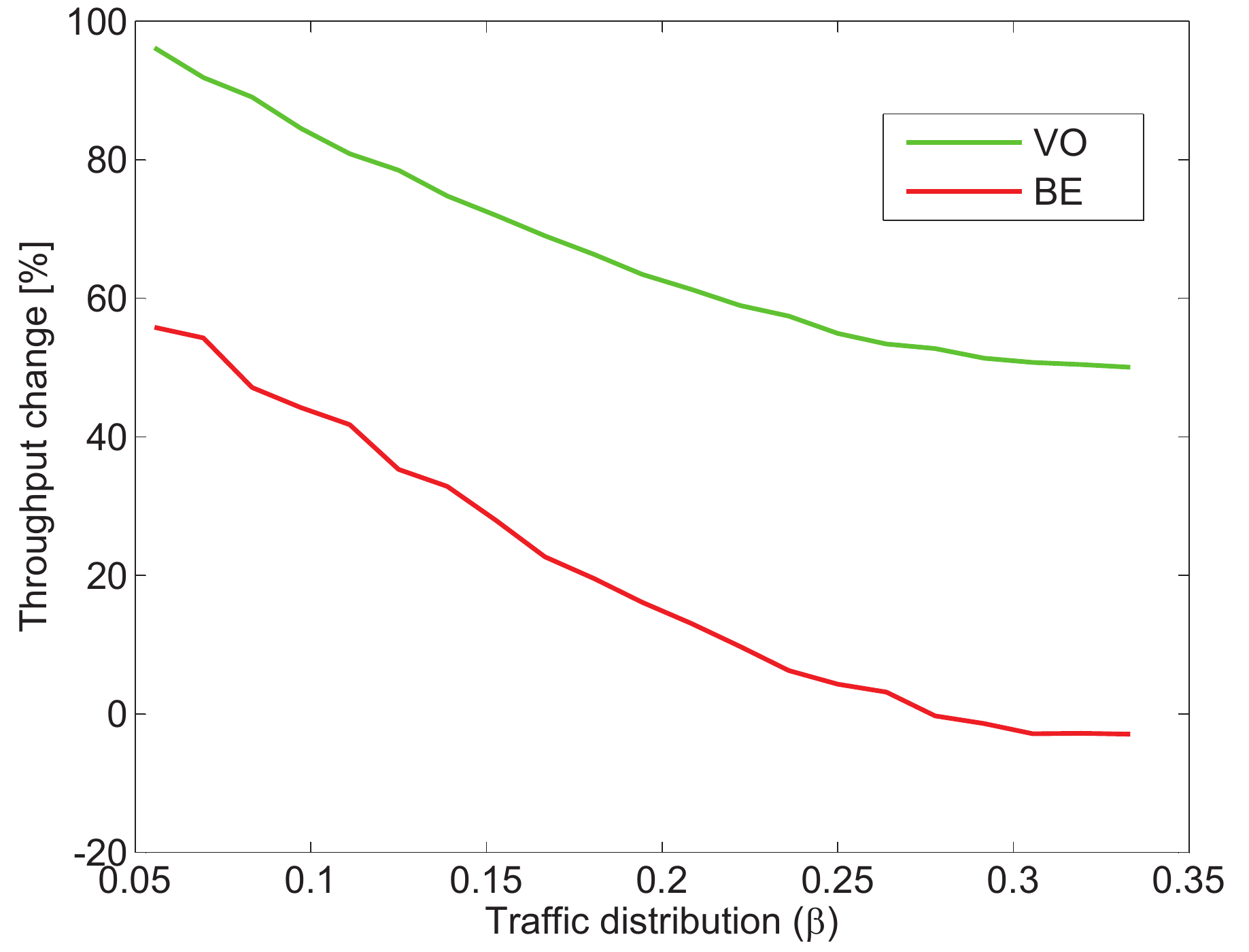}
		\caption{\scriptsize{DEMS average throughput change $T_{change}^{avg}[i](\beta)$ in comparison to 802.11ac}}
                 \label{VOBE_comp_3sta}
        \end{subfigure}
\caption{Comparison of the average throughput for DEMS and traditional scheduling as a function of $\beta$ for the three user case.
}
\end{figure*}

The comparison of the average VO and BE normalized throughput values as a function of $\beta$ 
for DEMS and 802.11ac are shown in Fig.~\ref{VO_comp_3sta} and Fig.~\ref{BE_comp_3sta}, respectively. They were calculated according to Eq.~(\ref{t_avg}). For DEMS the average VO and BE throughput values are stable for each $\beta$ and equal to 2.25 and 0.75, respectively. This means that on average all three users were receiving data simultaneously. For traditional scheduling both VO and BE's average throughput values are dependent on traffic distribution $\beta$. The minimum VO (BE) average throughput equals 1.14 (0.43) and the maximum VO (BE) average throughput equals 1.49 (0.69). Similarly to the two user case, this is a result of the impact of traffic uniformity on the HOL frame blocking probability.
Fig.~\ref{VOBE_comp_3sta}  additionally presents the average VO and BE DEMS throughput changes as a function of $\beta$ in comparison to 802.11ac, which were calculated according to Eq.~(\ref{t_ch_avg}). Similarly to the two user case, the average throughput change is positive for both VO and BE. For VO it ranges from $50\%$ to $96\%$ and for BE it ranges from $-3\%$ to $56\%$. Therefore, in comparison to the two user case, the average throughput improvement for the VO AC increases and for the BE AC decreases. 

\begin{figure*}[htb]
        \centering
        \begin{subfigure}[b]{0.4 \textwidth}
                \includegraphics[width=\textwidth]{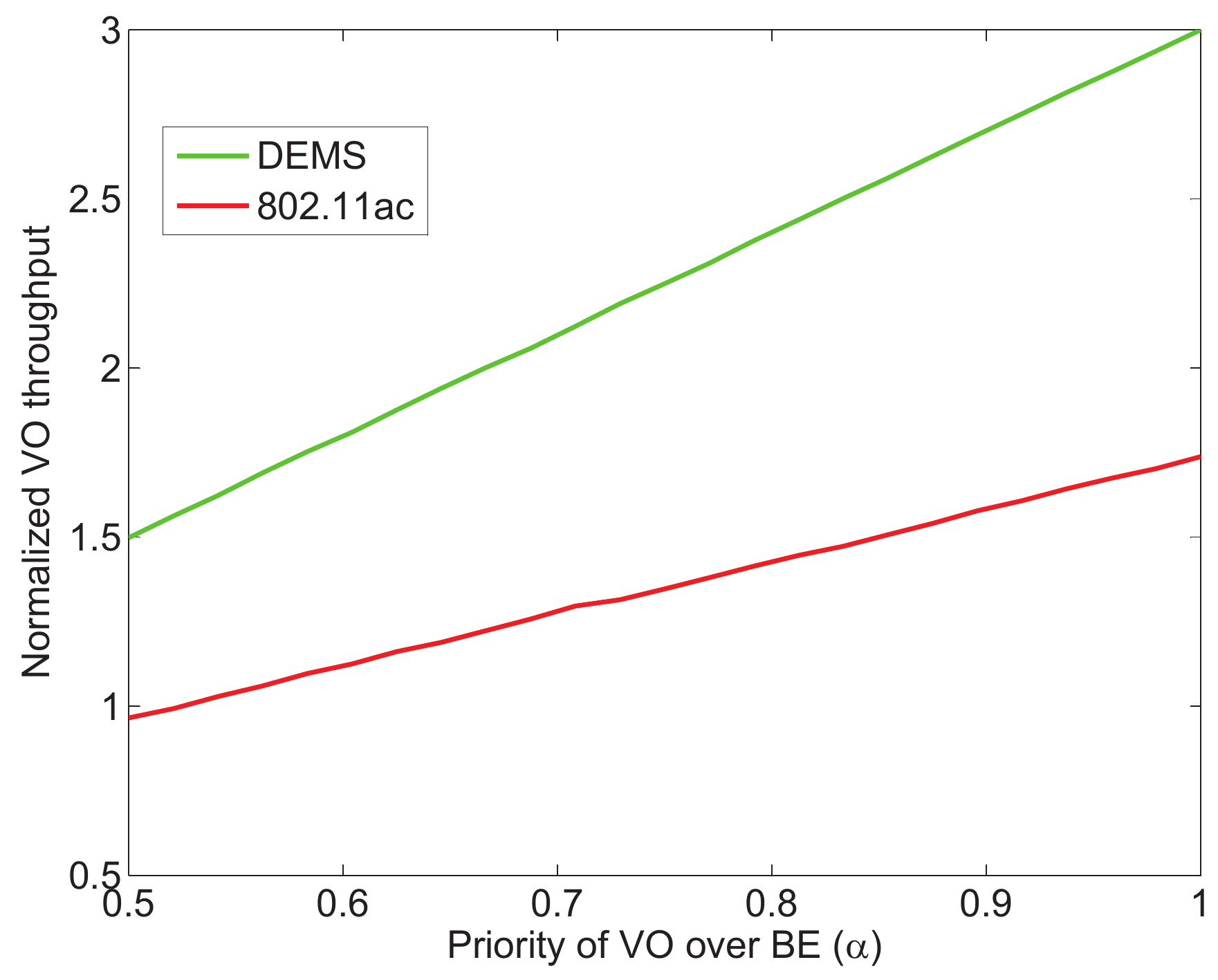}
                \caption{\scriptsize{Average VO throughput $T_{avg}[VO](\alpha)$}}
        \quad 
                \label{VO_comp_3sta-alpha}
        \end{subfigure}%
               \quad 
        \begin{subfigure}[b]{0.4 \textwidth}
                \includegraphics[width=\textwidth]{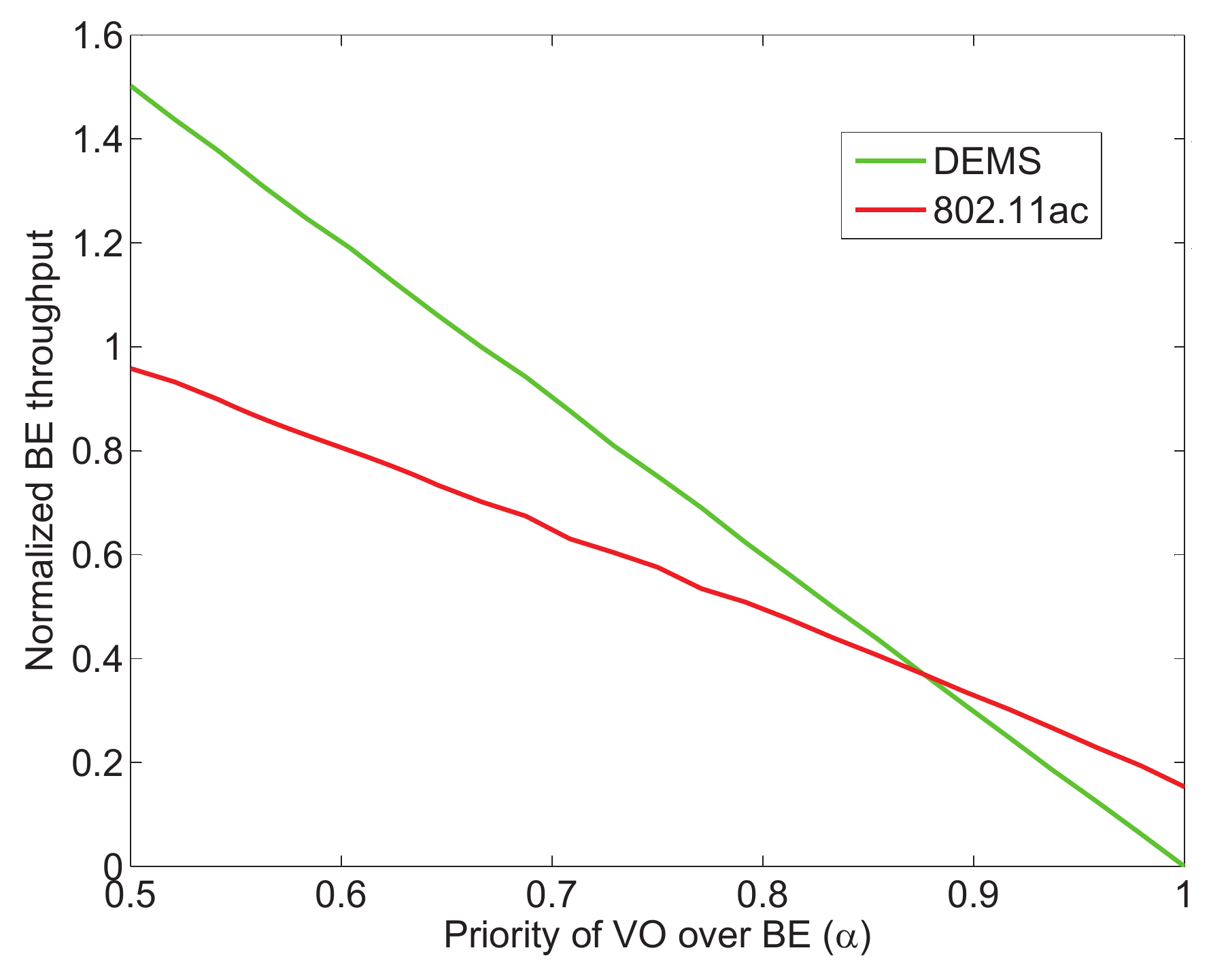}
		\caption{\scriptsize{Average BE throughput $T_{avg}[BE](\alpha)$}}
        \quad 
                 \label{BE_comp_3sta-alpha}
        \end{subfigure}
\quad
        \begin{subfigure}[b]{0.4 \textwidth}
                \includegraphics[width=\textwidth]{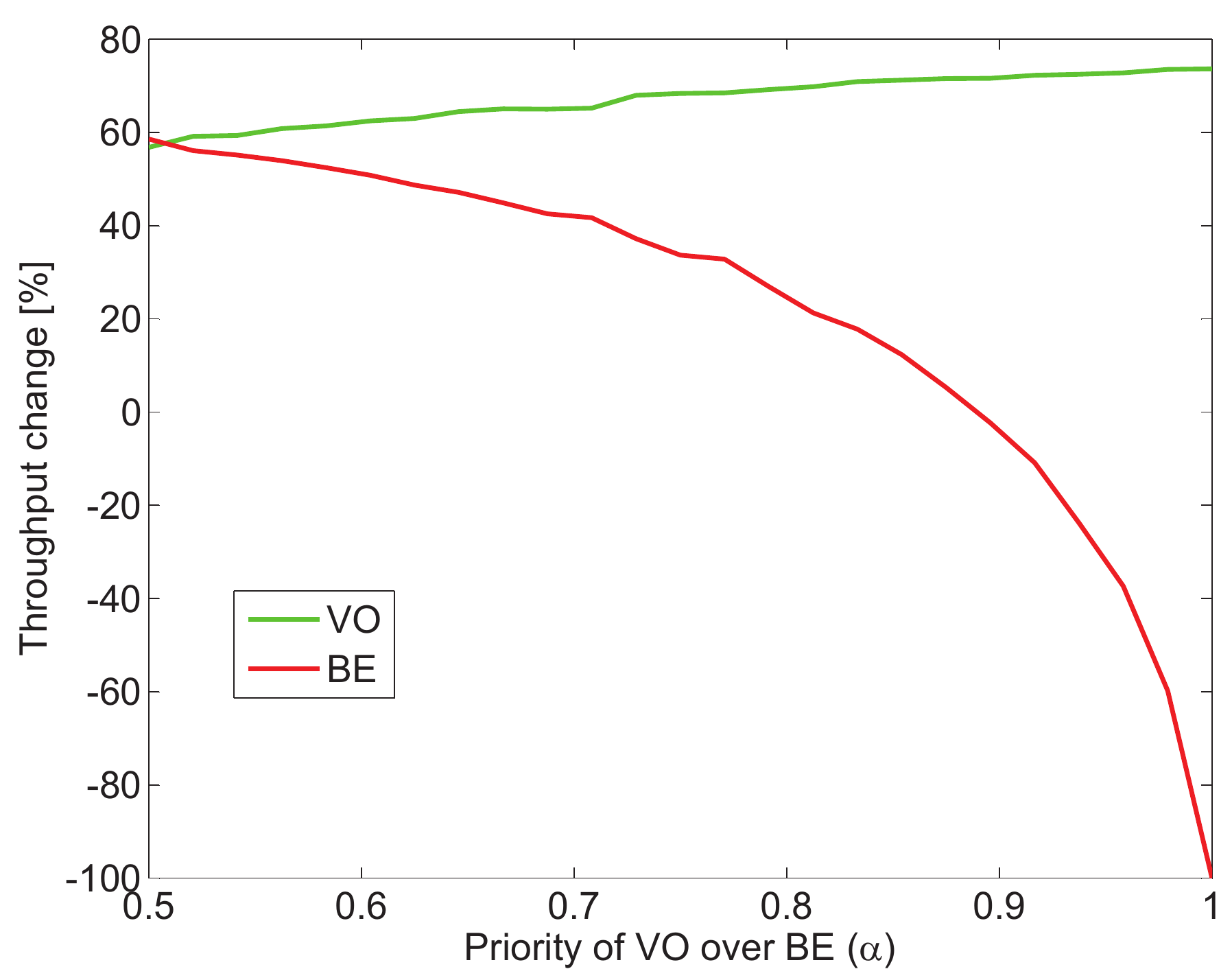}
		\caption{\scriptsize{DEMS average throughput change $T_{change}^{avg}[i](\alpha)$ in comparison to 802.11ac}}
                 \label{VOBE_comp_3sta-alpha}
        \end{subfigure}
\caption{Comparison of the average throughput for DEMS and traditional scheduling as a function of $\alpha$ for the three user case.
}
\end{figure*}

The comparison of the average VO and BE normalized throughput values as a function of $\alpha$ for DEMS and 802.11ac are shown in Fig.~\ref{VO_comp_3sta-alpha} and Fig.~\ref{BE_comp_3sta-alpha}, respectively. They were calculated according to Eq.~(\ref{t_avg-alpha}). For DEMS the average VO throughput increases from 1.5 to 3 and the average BE throughput decreases from 1.5 to 0 for increasing $\alpha$.  For 802.11ac the average VO throughput increases from 0.97 to 1.74 and the average BE throughput decreases from 0.96 to 0.15 for increasing $\alpha$.
Fig.~\ref{VOBE_comp_3sta-alpha}  additionally presents the average VO and BE DEMS throughput values as a function of $\alpha$ in comparison to 802.11ac, which were calculated according to Eq.~(\ref{t_ch_avg-alpha}). Similarly to the two user case, the average throughput change is always positive for VO (it ranges from $56.8\%$ to $73.6\%$) and for BE is decreases from $58.5\%$ to $-100\%$. Additionally, in comparison to the two user case, the average VO throughput improvement increases because the probability of HOL blocking in the traditional 802.11ac AP increases with the increasing number of the available spatial streams (cf. Fig.~\ref{hol}). As a final conclusion, DEMS preference over the traditional scheduling is expected for an increasing number of users involved in DL-MU transmissions (cf. 4- to 8-user cases in Fig.~\ref{hol}).

\section{Conclusions}
\label{concl}

In this paper we have proposed a new DEMS queuing mechanism for Wi-Fi networks supporting DL-MU-MIMO transmissions. It is based on the idea of decoupling EDCA and DL-MU-MIMO scheduling. Simulation results show that DEMS queuing outperforms traditional scheduling with FIFO queues. 

The most important advantages of DEMS queuing, in comparison to the traditional scheduling with FIFO queues, are the following:
improved throughput and decreased queuing delay (mostly for high priority ACs) because of better channel usage;
increased probability of transmission of time-critical traffic (e.g., video streaming) to end users, which could be advantageous, e.g., in cellular data offloading scenarios;
decreased competition between frames destined to different users with parallel processing replacing the traditional FIFO approach;
extendibility: the EDCA scheduling part can easily be replaced by, e.g, the 802.11aa intra-access category prioritization which provides better support for audio-video streaming \cite{prasnal1,prasnal2};
increased flexibility: by introducing two queue levels (class and transmission queues), one can either assign different ACs to the same user or the same AC to multiple users, therefore, more complex use cases can be considered.

As future work we envision further enhancements of the proposed mechanism, e.g., the definition of the per-frame scheduler, and its comparison with alternate solutions.

\begin{acknowledgements}
This work has been carried out as part of a project financed by the Polish National Science Centre (decision no. DEC-2011/01/D/ST7/05166). It was also supported by the Foundation for Polish Science (FNP).
The author would like to additionally thank PhD David Chieng and PhD Chien Su Fong from MIMOS and PhD Boris Bellalta from Universitat Pompeu Fabra for their valuable comments.
\end{acknowledgements}

\end{document}